# Tracing high energy radiation with molecular lines near deeply embedded protostars

P. Stäuber[1], A.O. Benz[1], J.K. Jørgensen[2], E.F. van Dishoeck[3], S.D. Doty[4], and F.F.S. van der Tak[5]

[1] Institute of Astronomy, ETH Zurich, CH-8092 Zurich, Switzerland
[2] Harvard-Smithsonian Center for Astrophysics, 60 Garden Street, Cambridge, MA 02138, USA
[3] Leiden Observatory, Leiden University, PO Box 9513, 2300 RA Leiden, The Netherlands
[4] Department of Physics and Astronomy, Denison University, Granville, OH 43023, USA
[5] SRON National Institute for Space Research, Landleven 12, 9747 AD Groningen, The Netherlands

**ABSTRACT**

*Aims.* The aim is to probe high energy radiation emitted by deeply embedded protostars.
*Methods.* Submillimeter lines of CN, NO, CO$^+$ and SO$^+$, and upper limits on SH$^+$ and N$_2$O are observed with the James Clerk Maxwell Telescope in two high-mass and up to nine low-mass young stellar objects and compared with chemical models.
*Results.* Constant fractional abundances derived from radiative transfer modeling of the line strengths are $x$(CN) ≈ a few ×$10^{-11}$–$10^{-8}$, $x$(NO) ≈ $10^{-9}$–$10^{-8}$ and $x$(CO$^+$) ≈ $10^{-12}$–$10^{-10}$. SO$^+$ has abundances of a few ×$10^{-11}$ in the high-mass objects and upper limits of ≈ $10^{-12}$–$10^{-11}$ in the low-mass sources. All abundances are up to 1–2 orders of magnitude higher if the molecular emission is assumed to originate mainly from the inner region ($\lesssim$ 1000 AU) of the envelope. For high-mass sources, the CN, SO$^+$ and CO$^+$ abundances and abundance ratios are best explained by an enhanced far-ultraviolet (FUV) field impacting gas at temperatures of a few hundred K. The observed column densities require that this region of enhanced FUV has scales comparable to the observing beam, such as in a geometry in which the enhanced FUV irradiates outflow walls. For low-mass sources, the required temperatures within the FUV models of $T \gtrsim 300$ K are much higher than found in models, so that an X-ray enhanced region close to the protostar ($r \lesssim 500$ AU) is more plausible. Gas-phase chemical models produce more NO than observed, suggesting an additional reduction mechanism not included in current models.
*Conclusions.* The observed CN, CO$^+$ and SO$^+$ abundances can be explained with either enhanced X-rays or FUV fields from the central source. High-mass sources likely have low opacity regions that allow the FUV photons to reach large distances from the central source. X-rays are suggested to be more effective than FUV fields in the low-mass sources. The observed abundances imply X-ray fluxes for the Class 0 objects of $L_X \approx 10^{29}$–$10^{31}$ erg s$^{-1}$, comparable to those observed from low-mass Class I protostars. Spatially resolved data are needed to clearly distinguish the effects of FUV and X-rays for individual species.

## 1. Introduction

The earliest phase of star-formation can only be probed through observations of molecular lines and dust continuum at (sub)millimeter and infrared wavelengths. Comparison of observations with detailed chemical models can put constraints on, for example, the ionization rate of the gas around young stellar objects (YSOs; e.g., Doty et al. 2004). Understanding the chemistry in these regions is therefore essential to gain knowledge of the physical processes involved at this stage.

The density and temperature distribution in envelopes around protostars can be derived by modeling the observed dust continuum. The chemical structure of the envelopes is constrained by molecular line observations, both in emission and absorption. To obtain molecular abundances, synthetic line fluxes are calculated with a radiative transfer model and compared to observations until agreement is found. This approach has been used successfully in high-mass (e.g., van der Tak et al. 2000; Boonman et al. 2003a) as well as in low-mass YSOs (e.g., Schöier et al. 2002; Jørgensen et al. 2004).

The density distribution can also be taken as a starting point for a full chemical model of the envelope. The molecular abundances are calculated in a large chemical network as a function of density, temperature and radial position from the central star. The models can then be used to determine the cosmic-ray ionization rate or to constrain the chemical time when compared to a large set of observations (e.g., Ceccarelli et al. 1996; Doty et al. 2002, 2004). In this way, Doty et al. (2004) found that an additional source of ionization was required for the low-mass Class 0 YSO IRAS 16293–2422 , since the observations could only be interpreted with an unusually high cosmic-ray ionization rate. It was shown by Stäuber et al. (2004, 2005) that FUV fields and X-rays from the central source provide supplementary ionizations and influence the chemistry of the inner envelope. These spherically symmetric envelope models generally improved the model fits to observations. FUV fields were found to affect only the chemistry in the innermost part of the envelope whereas X-rays penetrate deeper into the cloud due to the small cross sections at higher energies. X-rays can easily dominate over cosmic rays in terms of ionization rates in the inner part of the envelope.

It is still an open question, whether the youngest low-mass objects (Class 0 YSOs) emit X-rays (eg., Hamaguchi et al. 2005; Forbrich et al. 2006). High optical depths ($A_V > 150$) prevent possible X-rays from the young protostar to penetrate the surrounding envelope. X-ray luminosities observed towards more evolved objects are typically between $L_X \approx 10^{28}$–$10^{31}$ erg s$^{-1}$ in the 0.5–6 keV band with a thermal spectrum 0.6–7 keV (e.g., Imanishi et al. 2001; Preibisch et al. 2005). High X-ray luminosities with hard spectra are also observed towards high-mass YSOs (eg., Hofner et al. 2002; Townsley 2006). The X-ray luminosities typically observed towards low and high-mass sources are of the same order whereas the FUV fields from high-mass objects are expected to be much higher due to their higher stellar temperature. The amount of FUV photons emitted by young low-mass pre-main sequence stars is poorly known (e.g., Bergin et al. 2003). However, if the deeply embedded sources emit X-rays or FUV photons, the radiation might be traced by an enhanced chemistry. This is the goal of our investigation.

We present single-dish observations of molecular lines towards a sample of both low and high-mass YSOs (Sect. 2). The observed molecules are radicals and ions that are predicted to be enhanced by X-rays and FUV fields (Stäuber et al. 2004, 2005). To sample the dense inner region of the envelope rather than the outer part or large scale outflow material, the lines are chosen to have high critical densities ($n_{H_2} \approx 10^6$–$10^7$ cm$^{-3}$). The source sample is mainly based on the studies of Jørgensen et al. (2004) with focus on Class 0 objects for the low-mass sources and van der Tak et al. (2000) for the young high-mass objects. The FU Orionis source V1057 Cyg has been added to this list since this type of object shows surprisingly strong emission of ionized nitrogen at 122 $\mu$m which might be another indicator of energetic radiation (Lorenzetti et al. 2000). The observations are presented and discussed in Sect. 3. Radiative transfer is calculated to determine the molecular abundances in Sect. 4. To put constraints on the X-ray or FUV flux, gas density and temperature, the molecular abundances are calculated as a function of these parameters. The observations are compared to chemical models in Sect. 5. The results are discussed with focus on the inner source of radiation in Sect. 6. Other possible processes, such as shock chemistry, are not considered.

## 2. Observations

Heterodyne observations were carried out using the James Clerk Maxwell Telescope (JCMT) on Mauna Kea, Hawaii[1] between August 2003 and December 2005. The B3 receiver at 315–370 GHz was used with the digital autocorrelation spectrometer (DAS) in setups with bandwidths ranging from 125 MHz to 250 MHz. The main-beam efficiency was $\eta_{MB} = 0.63$ and the half-power beam width (HPBW) $\approx 14''$. The observed molecules and frequencies are listed in Table 1, the sources in Table 2. The CN lines as well as the two CO$^+$ lines were observed simultaneously in the upper and lower sideband, respectively. On source integration time was generally in the range of 1–3 hours and 10 hours for CO$^+$ in N1333–I2. The data were converted to the main-beam antenna temperature scale and analyzed with the CLASS software. The final spectra have a resolution of $\approx 0.27$–$0.55$ km s$^{-1}$ and rms noise levels between $\approx 11$–$50$ mK.

To derive the envelope structure of the FU Orionis object V1057 Cyg, continuum maps at 450 $\mu$m and 850 $\mu$m were obtained with the Submillimeter-User Bolometer Array (SCUBA) on the JCMT. The HPBW of SCUBA is $\approx 8''$ at 450 $\mu$m and $\approx 14''$ at 850 $\mu$m. The observations were performed in March 2004 with approximately one hour integration time.

## 3. Results

### 3.1. Line profiles

The observed spectra are presented in Figs. 1–6. CN was observed and detected in all sources listed in Table 2. CN was previously observed in the low-mass objects by Schöier et al. (2002) and Jørgensen et al. (2004). Current data, however, have higher signal to noise ratios to detect weaker hyperfine components that allow to determine the optical depth. NO was observed and detected in both high-mass sources as well as in N1333–I2 and N1333–I4A. CO$^+$ was observed towards all sources except L1489 but only detected in the high-mass sources, in IRAS 16293–2422 and tentatively in N1333–I2. SO$^+$ was observed in all sources except in L483 and L1489 but only detected in the high-mass sources and tentatively in IRAS 16293–2422. The line shapes are all fairly Gaussian with no or only weak outflow components (e.g., SMM4, Hogerheijde et al. 1999). Line intensities and line widths were calculated by fitting a Gaussian to each line. The results are presented in the Tables 3–5. The lines are relatively narrow and are not broadened due to, for example, inflow motion in the inner region (e.g., van der Tak et al. 2003) or indicative of a rotating disk, except for the case of L1489 (Hogerheijde 2001). For IRAS 16293–2422, self-absorption can affect the lines (e.g., van Dishoeck et al. 1995). The line fluxes in these cases are calculated by summing the individual components.

The narrow lines most likely trace the static envelope, either the bulk or a small part of it such as a geometrically thin

---

[1] The JCMT is operated by the Joint Astronomy Centre on behalf of the Particle Physics and Astronomy Research Council of the United Kingdom, the Netherlands Organization for Scientific Research, and the National Research Council of Canada.

**Table 2.** Sample of sources.

| Source | RA (J2000) (hh mm ss) | Dec (J2000) (dd mm ss) | $V_{LSR}$ (km/s) | $L_{bol}$ ($L_\odot$) | Distance (pc) | $M_{env}$ ($M_\odot$) | $r_{env}$ (AU) | $N(H_2)$ ($10^{23}$ cm$^{-2}$) | Type |
|---|---|---|---|---|---|---|---|---|---|
| AFGL 2591[a] | 20 29 24.7 | + 40 11 19 | -5.5 | $2 \times 10^4$ | 1000 | 44 | 27000 | 1.0 | HMPO |
| W3 IRS5[a] | 02 25 40.6 | + 62 05 51 | -39.0 | $1.7 \times 10^5$ | 2200 | 262 | 60000 | 6.6 | HMPO |
| IRAS 16293–2422[b] | 16 32 22.8 | − 24 28 33 | 4.0 | 27 | 160 | 5.4 | 8000 | 16.0 | Class 0 |
| N1333-I2[c] | 03 28 55.4 | + 31 14 35 | 7.0 | 16 | 220 | 1.7 | 21000 | 5.5 | Class 0 |
| N1333-I4A[c] | 03 29 10.3 | + 31 13 31 | 7.0 | 6 | 220 | 2.3 | 24000 | 22.0 | Class 0 |
| L1448-C[c] | 03 25 38.8 | + 30 44 05 | 5.0 | 5 | 220 | 0.93 | 15000 | 1.7 | Class 0 |
| L483[c] | 18 17 29.8 | − 04 39 38 | 5.2 | 9 | 200 | 1.1 | 32000 | 9.3 | Class 0 |
| L723[c] | 19 17 53.7 | + 19 12 20 | 11.2 | 3 | 300 | 0.62 | 21000 | 3.4 | Class 0 |
| SMM4[d] | 18 29 56.7 | + 01 13 15 | 7.9 | 11 | 250 | 5.8 | 9300 | 12.0 | Class 0 |
| V1057 Cyg[e] | 20 58 53.7 | + 44 15 28 | 4.1 | 70 | 650 | 1.4 | 15000 | 0.5 | FU Ori |
| L1489[c] | 04 04 43.0 | + 26 18 57 | 7.0 | 3.7 | 140 | 0.097 | 9400 | 1.0 | Class I |

The luminosity and envelope parameters are from [a] van der Tak et al. (1999, 2000); [b] Schöier et al. (2002); [c] Jørgensen et al. (2002); [d] Pontoppidan et al. (2004); [e] this paper. The $H_2$ column densities are calculated by integrating the power-law density distribution out to $r_{env}$ ($N(H_2) = \int n(r)\,dr$).

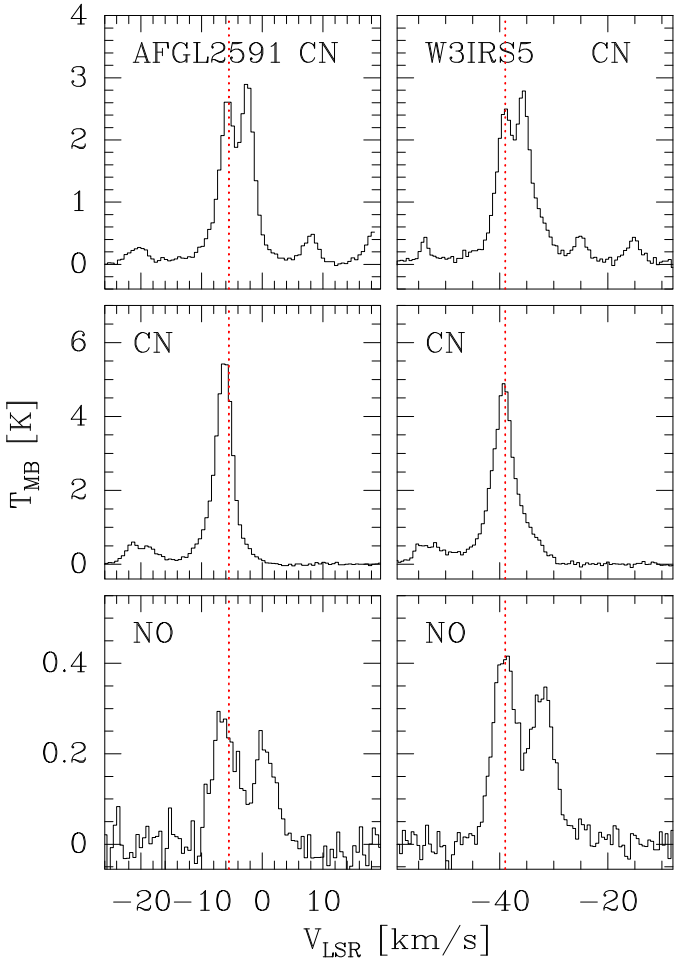

**Fig. 1.** CN and NO $J = 3 - 2$ transitions for the high-mass sources. The first row shows the CN $3\,\frac{5}{2}$ - $2\,\frac{3}{2}$ hyperfine transitions between $V_{LSR} \approx -10$ and $\approx 18$ km s$^{-1}$. The line at $V_{LSR} \approx -20$ km s$^{-1}$ is not identified. The CN $3\,\frac{7}{2}$ - $2\,\frac{5}{2}$ lines between $V_{LSR} \approx -22$ and $\approx -8$ km s$^{-1}$ are presented in the second row. The NO lines are in the third row. The dotted lines indicate the $V_{LSR}$ position of the sources.

**Table 1.** Transitions, frequencies, upper level energies ($E_{up}$) and Einstein $A$ coefficients ($A_{ul}$) of the observed lines.

| Molecule | Transition | Frequency (MHz) | $E_{up}$ (K) | $A_{ul}$ (s$^{-1}$) |
|---|---|---|---|---|
| CN | $3,\frac{5}{2},3 - 2,\frac{3}{2},3$ | 340008.1 | 32.6 | $6.2 \times 10^{-5}$ |
| CN | $3,\frac{5}{2},2 - 2,\frac{3}{2},2$ | 340019.6 | 32.6 | $9.3 \times 10^{-5}$ |
| CN | $3,\frac{5}{2},4 - 2,\frac{3}{2},3$ | 340031.5 | 32.6 | $3.9 \times 10^{-4}$ |
| CN | $3,\frac{5}{2},2 - 2,\frac{3}{2},1$ | 340035.4[a] | 32.6 | $2.9 \times 10^{-4}$ |
| CN | $3,\frac{7}{2},4 - 2,\frac{5}{2},3$ | 340247.8[b] | 32.7 | $3.8 \times 10^{-4}$ |
| CN | $3,\frac{7}{2},3 - 2,\frac{5}{2},2$ | 340248.6 | 32.7 | $3.7 \times 10^{-4}$ |
| CN | $3,\frac{7}{2},3 - 2,\frac{5}{2},3$ | 340261.8 | 32.7 | $4.5 \times 10^{-5}$ |
| NO | $\frac{7}{2},\frac{9}{2} - \frac{5}{2},\frac{7}{2}$ | 351043.5 | 36.1 | $5.4 \times 10^{-6}$ |
| NO | $\frac{7}{2},\frac{7}{2} - \frac{5}{2},\frac{5}{2}$ | 351051.7[c] | 36.1 | $5.0 \times 10^{-6}$ |
| $N_2O$ | 14 − 13 | 351667.8 | 126.6 | $6.3 \times 10^{-6}$ |
| $CO^+$ | $3,\frac{5}{2} - 2,\frac{3}{2}$ | 353741.3 | 33.9 | $1.6 \times 10^{-3}$ |
| $CO^+$ | $3,\frac{7}{2} - 2,\frac{5}{2}$ | 354014.2 | 34.0 | $1.7 \times 10^{-3}$ |
| $SO^+$ | $\frac{15}{2} - \frac{13}{2}$ | 347740.0 | 70.1 | $1.6 \times 10^{-3}$ |
| $SH^+$ | $1,\frac{3}{2} - 0,\frac{1}{2}$ | 345929.8 | 16.6 | $2.3 \times 10^{-4}$ |

[a] Overlap with hyperfine transition F = 3 − 2. [b] Overlap with hyperfine transition F = 5 − 4. [c] Overlap with hyperfine transition F = $\frac{5}{2}$ - $\frac{3}{2}$.

layer of gas covering the outflow cavity walls. Such a thin gas layer would, however, have to be centered at the cloud velocity, since the lines are not shifted from the local standard of rest velocity of the corresponding source. The influence of an outer FUV enhanced region can be excluded due to the high critical

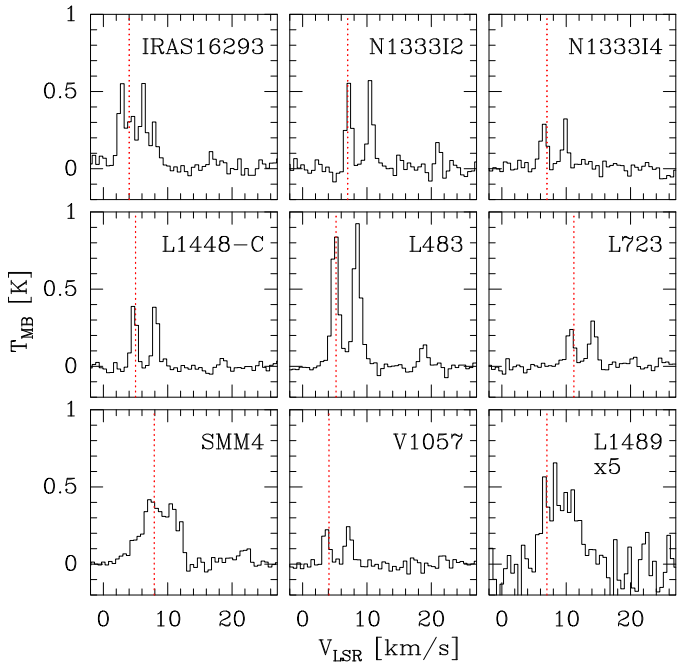

**Fig. 2.** CN 3 $\frac{5}{2}$ - 2 $\frac{3}{2}$ hyperfine lines for the low-mass sources. The spectrum of L1489 was multiplied by a factor of 5. The dotted lines indicate the $V_{\rm LSR}$ position of the sources.

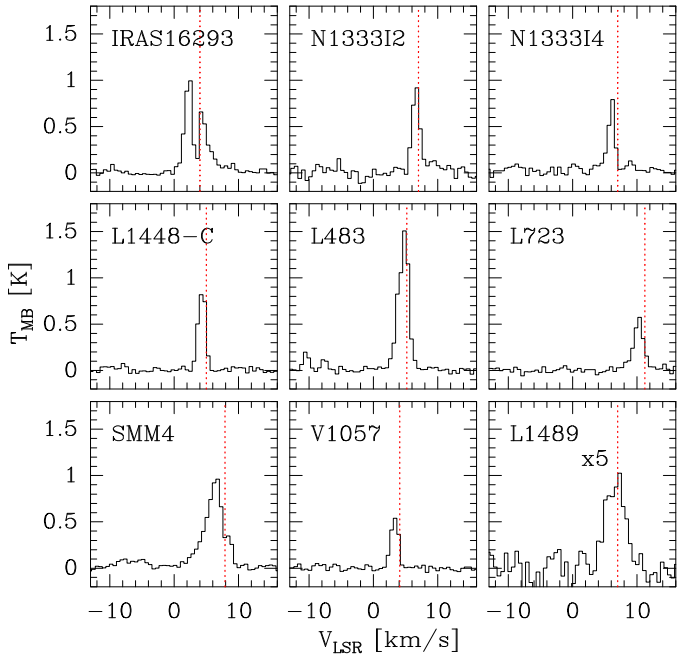

**Fig. 3.** CN 3 $\frac{7}{2}$ - 2 $\frac{5}{2}$ hyperfine lines for the low-mass sources. The spectrum of L1489 was multiplied by a factor of 5. The dotted lines indicate the $V_{\rm LSR}$ position of the sources.

densities of the observed transitions and previous model results by Stäuber et al. (2004, 2005).

In the case of non-detections, $3\sigma$ upper limits are given ($1\sigma = 1.2\sqrt{\delta V \Delta V}\sigma_{\rm rms}$, where $\Delta V$ is the expected line width, $\delta V$ the channel width and $\sigma_{\rm rms}$ is the rms noise in the observed spectra, Jørgensen et al. 2004). The calibration uncertainty is taken to be 20%. $N_2O$ at 351.668 GHz was observed in W3

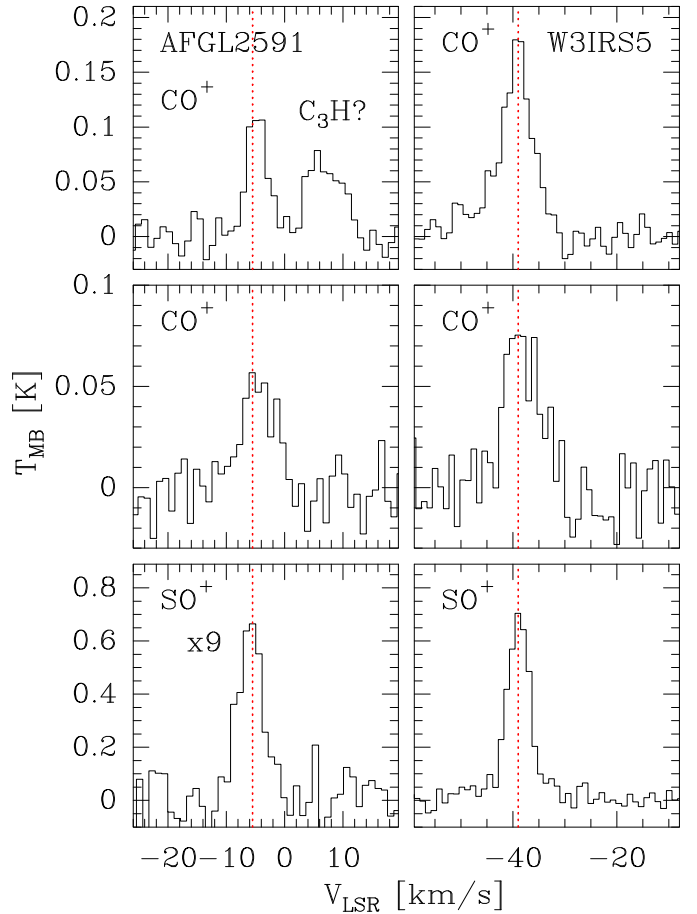

**Fig. 4.** $CO^+$ and $SO^+$ $J = 3 - 2$ transitions for the high-mass sources. The $CO^+$ $J = 3\,\frac{5}{2} - 2\,\frac{3}{2}$ lines are presented in the first row. The $CO^+$ $J = 3\,\frac{7}{2} - 2\,\frac{5}{2}$ transitions are shown in the second row. $SO^+$ is given in the third row. The $SO^+$ spectrum of AFGL 2591 was multiplied by a factor of 9. The dotted lines indicate the $V_{\rm LSR}$ position of the sources.

IRS5 and N1333–I2 but not detected. A $3\sigma$ upper limit of 0.06 K km s$^{-1}$ is derived for W3 IRS5 and N1333–I2. $SH^+$ was searched for in AFGL 2591 at 345.930 GHz. Unfortunately the line is blended with a $^{34}SO_2$ line at 345.929 GHz. The upper limit derived for $SH^+$ will be discussed in Sect. 3.1.3. Judging from Tables 3–5, V1057 Cyg does not appear to be unusual in its chemistry in spite of its strong [NII] emission.

### 3.1.1. CN and NO

The CN lines in the high-mass sources (Fig. 1) look similar with almost identical main-beam temperatures and line widths (Tables 3 and 4). The strongest CN lines in W3 IRS5 show wings which indicate that some CN is at a different velocity, probably due to outflows.

The CN lines of the low-mass sources are presented in Figs. 2 and 3. The strongest CN lines among the low-mass sources are found in L483. The weakest lines are those of the Class I object L1489. The NO main-beam temperatures are $\approx 10\times$ lower than those of the strongest CN lines. The lines are presented in Fig. 1 for the high-mass objects and in Fig. 6 for the two N1333 sources.

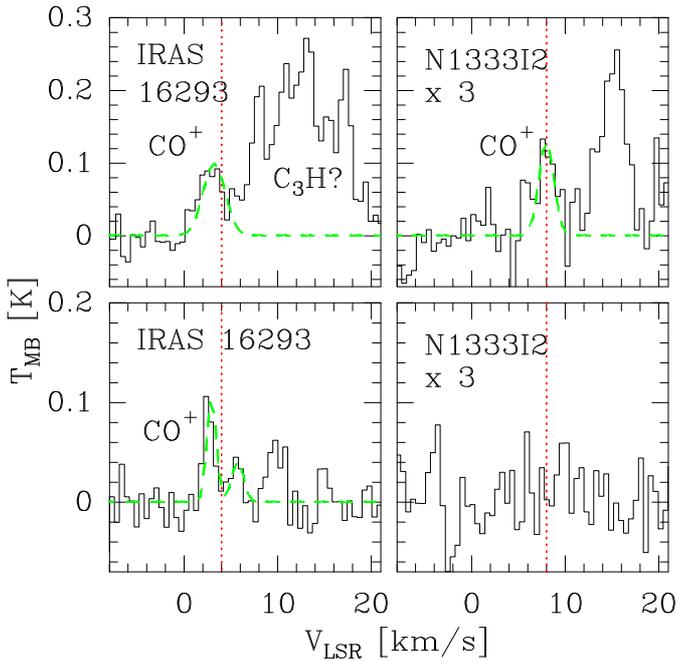

**Fig. 5.** $CO^+$ IRAS 16293–2422 and N1333–I2. The N1333–I2 spectra are multiplied by a factor of 3. The $CO^+$ $J = 3\frac{5}{2} - 2\frac{3}{2}$ lines are presented in the first row. The $CO^+$ $J = 3\frac{7}{2} - 2\frac{5}{2}$ transitions are shown in the second row. The features seen to the right of the $CO^+$ lines are unidentified lines (see text). The dotted lines indicate the $V_{LSR}$ position of the sources.

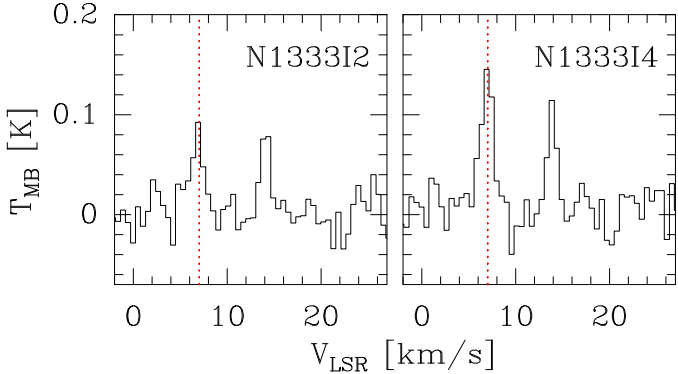

**Fig. 6.** NO $J = 3 - 2$ transitions for the N1333 low-mass sources. The dotted lines indicate the $V_{LSR}$ position of the sources.

The hyperfine structure of the CN and NO lines allows to calculate the optical depth of these lines with the HFS method described within CLASS. The method assumes the same excitation temperature and line width for all components of the multiplet. The derived optical depths $\tau$ of CN are given in Tables 3 and 4, those of NO are presented in Table 5. The calculated errors are typically between 5–20%. None of the lines appear to be highly optically thick since $\tau < 1$ for all lines.

### 3.1.2. $CO^+$ and $SO^+$

$CO^+$ is detected for the first time towards AFGL 2591, W3 IRS5 and tentatively N1333–I2. Previously, $CO^+$ has been observed towards PDRs, planetary nebulae (e.g., Latter et al. 1993; Fuente et al. 2003; Savage & Ziurys 2004) and tentatively towards the low-mass YSO IRAS16293–2422 (Ceccarelli et al. 1998). The $CO^+$ line at 353.741 GHz detected towards IRAS16293–2422 is surprisingly strong and has a similar peak temperature as the lines in the high-mass objects. The detection of $CO^+$ at 353.741 GHz towards N1333–I2 is at the $3\sigma$ level with a main-beam temperature $T_{MB} = 0.04$ K. The line flux is even at $6\sigma$ (Table 5). However, the 354.014 GHz line is not seen in the spectrum and $CO^+$ is thus only tentatively detected towards N1333–I2. The 353.7/354.0 line ratio is estimated to be $\approx 0.8 - 1.2$ from radiative transfer calculations. The line at 353.7 GHz appears to be $\approx 2-3$ times stronger in all observations, which could be due to different excitation mechanisms (Appendix A). Another possibility is that the line at 353.741 GHz is blended with an unidentified line. The $CO^+$ spectra at 353.741 GHz of AFGL 2591, IRAS 16293–2422 and N1333–I2 show one or several lines at $\approx 353.730$ GHz. The lines could be due to emission of $C_3H$ at 353.731 GHz and 353.733 GHz and $CH_3OOH$ at 353.736 GHz. These features, however, are not seen in the spectrum of W3 IRS5, where the $CO^+$ 353.7/354.0 line ratio is also $\approx 2$. In addition, the observed sources are not extremely rich in complex molecules (no line confusion; e.g., Helmich & van Dishoeck 1997). The lines are therefore assumed to be real in the following sections.

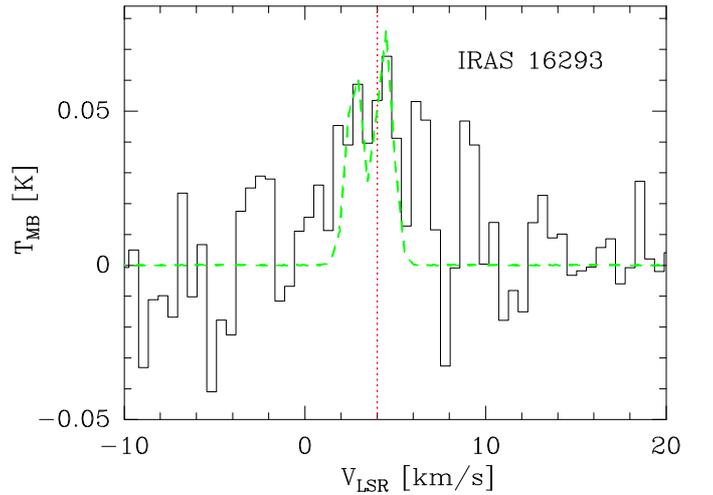

**Fig. 7.** Spectrum at 347.740 GHz of $SO^+$ towards IRAS 16293–2422. The dotted line indicates the $V_{LSR}$ position of the sources.

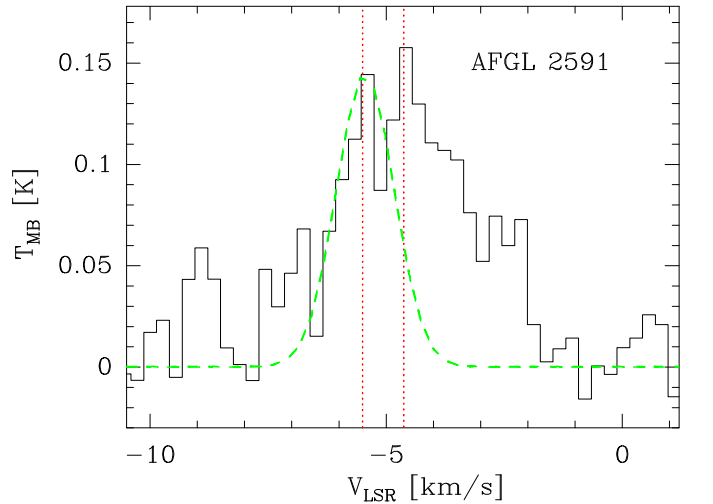

**Fig. 8.** Spectrum at 345.930 GHz. The two dotted vertical lines indicate the frequencies 345.930 GHz ($SH^+$) and 345.929 GHz ($^{34}SO_2$), respectively, at the $V_{LSR}$ of the source. The spectrum is overlaid with a Gaussian fit to be viewed as an upper limit for $SH^+$.

The $SO^+$ line in W3 IRS5 is $\approx 9$ times as strong as that in AFGL 2591 (Fig. 4). This is most probably due to the unusually high sulphur abundance found in W3 IRS5 (e.g., van der Tak et al. 2003). When $SO^+$ was first detected towards the supernova remnant IC443 (Turner 1992), it was proposed as a tracer of dissociative shocks. When later surveys, however, showed large amounts of $SO^+$ towards dark clouds and star forming regions, it was suggested that $SO^+$ is not associated to shocks (Turner 1994). Indeed, the line profiles of the observed $SO^+$ lines do not indicate signs of shocked gas (Fig. 4). In the low-mass sample, $SO^+$ is tentatively detected towards IRAS 16293–2422 (Fig. 7). Assuming the lines to be doubly peaked as is the case for CN and $CO^+$, the integrated $SO^+$ line fluxes are 0.08 K km s$^{-1}$ and 0.09 K km s$^{-1}$, corresponding to $\approx 4\sigma$ detections. Nevertheless, the lines are treated as upper limits due to the noisy spectrum (Table 5).

### 3.1.3. $SH^+$

The $N = 1\frac{3}{2} - 0\frac{1}{2}$ line of $SH^+$ was searched for at 345.930 GHz (Savage et al. 2004) towards AFGL 2591. Unfortunately, the line is blended with the $^{34}SO_2$ ($J = 19_{1,18} - 18_{0,18}$) line at 345.929 GHz. Van der Tak et al. (2003) observed line fluxes between 1.0–1.2 K km s$^{-1}$ for $^{34}SO_2$ with line widths between 3.5–4.5 km s$^{-1}$. A Gaussian fit of the observed $^{34}SO_2$ line at 345.930 GHz leads to a line width $\Delta V = 3.8$ km s$^{-1}$ and an integrated flux of 0.5 K km s$^{-1}$. The values of the line flux and width are well in the range of those observed for other $^{34}SO_2$ transitions by van der Tak et al. (2003). Motivated by the fact that the two lines seen in Fig. 8 are exactly at 345.930 GHz and 345.929 GHz, an upper limit for the line flux of $SH^+$ can be obtained by assuming the lines to be $SH^+$ and $^{34}SO_2$, respectively. The width of the $SH^+$ line is assumed to be 1.5 km s$^{-1}$. This might be too small but larger widths lead to poorer line fits. The Gaussian fit gives an integrated intensity of 0.23 K km s$^{-1}$ for $SH^+$ as an upper limit for AFGL 2591. It should be mentioned though, that $SH^+$ has never been observed in the interstellar medium. A clear identification of the molecule therefore requires detections of at least two different transitions. This might be possible with the future Herschel Space Observatory, since the higher lying lines of $SH^+$ are not detectable with ground based telescopes (Savage et al. 2004).

### 3.2. Continuum observations of V1057

The peak flux densities observed towards V1057 with SCUBA are 2.0 and 0.35 Jy beam$^{-1}$ at 450 and 850 $\mu$m, respectively. The result for 850 $\mu$m is comparable to the observations of Sandell & Weintraub (2001), who studied this source in detail.

V1057 was not included in the source sample of Jørgensen et al. (2002). A model is thus constructed for the envelope using the same approach reproducing SCUBA maps at 450 $\mu$m and 850 $\mu$m and the SED from 25 $\mu$m–850 $\mu$m from the literature (in particular, ISOPHOT measurements of Abraham et al. 2004). The V1057 envelope is found to be well-fit by a power-law density profile $n \propto r^{-p}$ with $p = 1.3$ and $n_{H_2}(1000 AU) = 2.9 \times 10^5$ cm$^{-3}$ for radii of 10 AU to 15000 AU and for a total luminosity of 70 $L_\odot$ (at a distance of 650 pc). Previously Kenyon & Hartmann (1991) modeled the SED of V1057 focusing primarily on the IRAS measurements; their results for the V1057 envelope are found to be similar to ours. The luminosity, envelope mass, $H_2$ column density and approximate size of the envelope (radius) are presented in Table 2.

## 4. Molecular abundances

### 4.1. Constant abundance models

In order to estimate molecular abundances, a radiative transfer analysis is performed by using the 1D Monte Carlo radiative transfer code by Hogerheijde & van der Tak (2000). The program solves for the molecular excitation as a function of radius. The principal input parameters of the code are the radial $H_2$ density and gas temperature distribution, which were derived from dust radiative transfer analysis described in Jørgensen et al. (2002), assuming $T_{gas} = T_{dust}$. For the high-mass sources we use the results of van der Tak et al. (1999, 2000), for IRAS 16293–2422, the results of Schöier et al. (2002), for SMM4 we follow Pontoppidan et al. (2004), the results for V1057 Cyg are taken from Sect. 3.2. The parameters for all other low-mass sources are used from the studies of Jørgensen et al. (2002). The molecular abundances for the species of interest are assumed to be constant with radius. The result is integrated over the line of sight and convolved with a 14″ telescope beam. Observed and synthetic line fluxes are then compared with a $\chi^2$ statistic to find the best-fit abundance. The individual hyperfine components of CN are modeled as separate lines and the integrated line fluxes are summed in case of overlapping lines (eg., CN $3\frac{7}{2} - 2\frac{5}{2}$ and CN $3\frac{9}{2} - 2\frac{7}{2}$). The collisional rate coefficients for CN are obtained by scaling the coefficients of CO (Green & Chapman 1978). Estimates by Black (2004, private communication) are used for the $CO^+$ coefficients. For molecules with unknown collision rates (NO, $N_2O$, $SO^+$ and $SH^+$), the excitation is assumed to be thermalized at the temperature of each grid point of the model.

Table 6 lists the inferred abundances for CN, NO, $CO^+$ and $SO^+$. The upper limits for $CO^+$ and $SO^+$ correspond to $3\sigma$ in line flux (Table 5). Our CN abundances are consistent within a factor of two to those found by Jørgensen et al. (2004) and Schöier et al. (2002) for the same sources. CN was also observed by Helmich & van Dishoeck (1997) towards W3 IRS5. They derived a fractional abundance of $6.9 \times 10^{-10}$ for CN, which is $\approx 5$ times lower than ours. This is most probably due to the different methods used to derive the molecular abundance (full physical modeling vs. beam-averaged column density ratio). The reduced $\chi^2$ values from 3 to 7 different CN transitions per source (see Tables 3 and 4) are $\lesssim 1$, indicating that the

**Table 3.** Observed line fluxes ($W = \int T_{MB}\, dV$ in K km s$^{-1}$), line widths ($\Delta V$ in km s$^{-1}$) and opacities for the CN $3\,\frac{5}{2} - 2\,\frac{3}{2}$ transitions. The frequencies are given in GHz. Upper limits are at $3\sigma$ in line flux. Numbers in brackets are uncertainties in units of the the last decimal place. The line width was taken to be fixed with an estimated error of $\lesssim 30\%$.

| Source | 340.035 | | | 340.031 | | | 340.019 | | | 340.008 | | |
|---|---|---|---|---|---|---|---|---|---|---|---|---|
| | W | ΔV | τ | W | ΔV | τ | W | ΔV | τ | W | ΔV | τ |
| AFGL 2591 | 7.71(14) | 2.7 | 0.27(3) | 8.39(14) | 2.7 | 0.29(3) | 1.40(13) | 2.7 | 0.05(1) | 1.49(20) | 2.7 | 0.05(1) |
| W3 IRS 5 | 8.34(22) | 3.1 | 0.25(5) | 9.04(22) | 3.1 | 0.27(5) | 1.52(20) | 3.1 | 0.05(1) | 1.36(20) | 3.1 | 0.04(1) |
| IRAS 16293 | 1.14(7) | 1.2 | 0.52(5) | 0.92(6) | 1.0 | 0.53(5) | 0.12(5) | 1.0 | 0.10(1) | <0.06 | ... | ... |
| N1333-I2 | 0.61(4) | 1.0 | 0.06(1) | 0.59(4) | 1.0 | 0.06(1) | 0.16(4) | 0.8 | 0.02(1) | <0.10 | ... | ... |
| N1333-I4A | 0.38(4) | 1.2 | 0.09(1) | 0.31(4) | 0.9 | 0.08(1) | <0.09 | ... | ... | <0.09 | ... | ... |
| L1448-C | 0.46(2) | 1.1 | 0.12(1) | 0.45(2) | 1.1 | 0.12(1) | 0.06(2) | 1.1 | 0.02(<1) | 0.06(2) | 1.1 | 0.01(<1) |
| L483 | 1.21(10) | 1.3 | 0.09(1) | 1.28(10) | 1.3 | 0.09(1) | 0.19(10) | 1.2 | 0.01 | 0.20(10) | 1.0 | 0.02(<1) |
| L723 | 0.36(3) | 1.4 | 0.03(<1) | 0.40(3) | 1.3 | 0.03(<1) | <0.06 | ... | ... | <0.06 | ... | ... |
| SMM4 | 1.36(4) | 3.0 | 0.05(2) | 0.98(4) | 2.5 | 0.05(2) | 0.22(5) | 2.2 | 0.01(<1) | 0.16(4) | 2.0 | 0.01(<1) |
| V1057 Cyg | 0.29(3) | 1.1 | 0.11(2) | 0.31(3) | 1.2 | 0.11(2) | <0.07 | ... | ... | <0.07 | ... | ... |
| L1489 | 0.28(5) | 1.1 | 0.01(<1) | 0.21(5) | 1.1 | 0.01(<1) | <0.07 | ... | ... | <0.07 | ... | ... |

**Table 4.** Observed line fluxes ($W = \int T_{MB}\, dV$ in K km s$^{-1}$), line widths ($\Delta V$ in km s$^{-1}$) and opacities for the CN $3\,\frac{7}{2} - 2\,\frac{5}{2}$ transitions. The frequencies are given in GHz. Upper limits are at $3\sigma$ in line flux. Numbers in brackets are uncertainties in units of the the last decimal place. The hyperfine line widths were taken to be fixed with an estimated error of $\lesssim 30\%$.

| Source | 340.265 | | | 340.261 | | | 340.247 | | |
|---|---|---|---|---|---|---|---|---|---|
| | W | ΔV | τ | W | ΔV | τ | W | ΔV | τ |
| AFGL 2591 | 1.59(14) | 2.6 | 0.06(1) | 1.36(14) | 2.7 | 0.05(1) | 19.21(15) | 3.3 | 0.55(1) |
| W3 IRS 5 | 1.66(33) | 2.6 | 0.06(1) | 1.48(13) | 2.5 | 0.06(1) | 20.68(37) | 3.8 | 0.51(1) |
| IRAS 16293 | <0.06 | ... | ... | 0.08(2) | 0.8 | 0.02(1) | 2.01(11) | 1.3 | 0.26(2) |
| N1333-I2 | 0.10(5) | 0.9 | 0.01(<1) | 0.11(5) | 0.7 | 0.02(1) | 1.29(5) | 1.2 | 0.10(1) |
| N1333-I4A | <0.09 | ... | ... | <0.09 | ... | ... | 1.01(6) | 1.2 | ... |
| L1448-C | <0.06 | ... | ... | 0.09(7) | 1.1 | 0.01(<1) | 1.28(7) | 1.4 | 0.12(2) |
| L483 | 0.24(3) | 1.1 | 0.02(1) | 0.15(2) | 1.1 | 0.01(<1) | 2.78(3) | 1.7 | 0.15(9) |
| L723 | <0.06 | ... | ... | <0.06 | ... | ... | 1.03(3) | 1.7 | ... |
| SMM4 | 0.23(4) | 2.2 | 0.01(<1) | 0.22(4) | 2.2 | 0.01(<1) | 2.86(4) | 2.8 | 0.10(2) |
| V1057 Cyg | <0.07 | ... | ... | <0.07 | ... | ... | 0.85(3) | 1.4 | ... |
| L1489 | <0.07 | ... | ... | <0.07 | ... | ... | 0.72(5) | 3.6 | ... |

**Table 5.** Observed line fluxes ($W = \int T_{MB}\, dV$ in K km s$^{-1}$) and line widths ($\Delta V$ in km s$^{-1}$) for the NO, CO$^+$ and SO$^+$ transitions. The frequencies are given in GHz. Upper limits are at $3\sigma$ in line flux. Numbers in brackets are uncertainties in units of the the last decimal place. The hyperfine line widths of NO were taken to be fixed with an estimated error of $\lesssim 30\%$.

| Source | NO 351.043 | | | NO 351.051 | | | CO$^+$ 353.741 | | CO$^+$ 354.014 | | SO$^+$ 347.740 | |
|---|---|---|---|---|---|---|---|---|---|---|---|---|
| | W | ΔV | τ | W | ΔV | τ | W | ΔV | W | ΔV | W | ΔV |
| AFGL 2591 | 0.99(6) | 4.1 | 0.28(2) | 1.30(6) | 4.1 | 0.21(1) | 0.50(6) | 4.0(6) | 0.27(3) | 4.0[a] | 0.40(3) | 5.1(5) |
| W3 IRS 5 | 1.79(5) | 4.7 | 0.37(9) | 2.21(5) | 4.7 | 0.30(7) | 0.82(4) | 3.6(5) | 0.37(4) | 3.6[a] | 3.59(10) | 4.5(6) |
| IRAS 16293 | ... | ... | ... | ... | ... | ... | 0.31(4) | 2.9[a] | 0.21(7) | 2.7(6) | ≲0.17 | ... |
| N1333-I2 | 0.10(2) | 0.9 | 0.01(1) | 0.10(2) | 0.9 | 0.01(1) | ≲0.09 | 1.9 | <0.05 | ... | <0.04 | ... |
| N1333-I4A | 0.13(2) | 0.9 | 0.01(1) | 0.24(2) | 1.5 | 0.01(1) | <0.06 | ... | <0.06 | ... | <0.04 | ... |
| L1448-C | ... | ... | ... | ... | ... | ... | <0.05 | ... | <0.05 | ... | <0.04 | ... |
| L483 | ... | ... | ... | ... | ... | ... | <0.04 | ... | <0.04 | ... | ... | ... |
| L723 | ... | ... | ... | ... | ... | ... | <0.05 | ... | <0.05 | ... | <0.07 | ... |
| SMM4 | ... | ... | ... | ... | ... | ... | <0.05 | ... | <0.05 | ... | <0.07 | ... |
| V1057 Cyg | ... | ... | ... | ... | ... | ... | <0.05 | ... | <0.05 | ... | <0.07 | ... |

[a] The line width was taken to be fixed due to the noisy spectrum.

**Table 6.** Inferred abundances from radiative transfer models assuming constant radial abundances.

| Source | CN | NO | CO$^+$ | SO$^+$ |
|---|---|---|---|---|
| AFGL 2591 | $2.3 \times 10^{-08}$ | $3.1 \times 10^{-08}$ | $1.6 \times 10^{-10}$ | $5.8 \times 10^{-11}$ |
| W3 IRS 5 | $3.6 \times 10^{-09}$ | $1.7 \times 10^{-08}$ | $3.9 \times 10^{-11}$ | $7.0 \times 10^{-11}$ |
| IRAS 16293 | $5.4 \times 10^{-11}$ | ... | $1.5 \times 10^{-12}$ | $\lesssim 3.0 \times 10^{-12}$ |
| N1333-I2 | $3.9 \times 10^{-10}$ | $2.5 \times 10^{-09}$ | $\lesssim 7.5 \times 10^{-12}$ | $< 5.0 \times 10^{-12}$ |
| N1333-I4A | $4.2 \times 10^{-11}$ | $1.5 \times 10^{-09}$ | $< 1.3 \times 10^{-12}$ | $< 2.2 \times 10^{-12}$ |
| L1448-C | $1.1 \times 10^{-09}$ | ... | $< 2.0 \times 10^{-11}$ | $< 1.4 \times 10^{-11}$ |
| L483 | $1.5 \times 10^{-09}$ | ... | $< 7.0 \times 10^{-12}$ | ... |
| L723 | $9.2 \times 10^{-10}$ | ... | $< 2.0 \times 10^{-11}$ | $< 4.3 \times 10^{-11}$ |
| SMM4 | $1.6 \times 10^{-10}$ | ... | $< 1.0 \times 10^{-12}$ | $< 4.0 \times 10^{-12}$ |
| V1057 Cyg | $5.4 \times 10^{-09}$ | ... | $< 2.0 \times 10^{-10}$ | $< 5.0 \times 10^{-11}$ |
| L1489 | $2.5 \times 10^{-09}$ | ... | ... | ... |

constant abundance models fit the observations well. The same is true for NO. In the case of CO$^+$, $\chi^2$ is between 1–2 for all sources. However, the number of lines are two at most for CO$^+$ and NO, providing rather poor statistics.

NO was first observed by Liszt & Turner (1978) towards the molecular cloud Sgr B2. They reported a fractional abundance of $\approx 10^{-8}$. Similar abundances were derived by Ziurys et al. (1991) towards other star-forming clouds. Our observed NO abundances towards the high-mass and the N1333 sources (Table 6) are comparable to these values.

For N$_2$O, an upper limit of $x(N_2O) \lesssim 5 \times 10^{-10}$ is derived for W3 IRS5 and $x(N_2O) \lesssim 7 \times 10^{-9}$ for N1333–I2. The relatively high upper limits are due to the rather low Einstein spontaneous transition coefficient for the observed emission (Table 1). The upper limit for SH$^+$ in AFGL 2591 is estimated to be $x(SH^+) \lesssim 4.1 \times 10^{-11}$.

SO$^+$ in W3 IRS5 was observed by Turner (1994) and Helmich & van Dishoeck (1997). Our derived fractional SO$^+$ abundance differs by only a factor of 2.7 compared to the results of Turner (1994). No SO$^+$ abundances were reported by Helmich & van Dishoeck (1997). Their line strength, however, is in good agreement to our observation. Ceccarelli et al. (1998) inferred a CO$^+$ column density $N(CO^+) \approx 10^{11}$–$10^{12}$ cm$^{-2}$ - depending on the size of the CO$^+$ emitting region - from spectrally unresolved ISO observations towards IRAS 16293–2422 in a 80$''$ beam. Our observed column density toward the low-mass source IRAS 16293–2422 is $N(CO^+) \approx 10^{12}$ cm$^{-2}$.

Fractional abundances of CO$^+$ and SO$^+$ observed toward photon-dominated regions (PDRs) are usually a few times $10^{-11}$–$10^{-10}$ with column densities between $10^{12}$–$10^{13}$ cm$^{-2}$ (Fuente et al. 2003). Our abundances observed towards the high-mass sources are thus of the order of those found towards PDRs, indicating that FUV fields from the central source may be important.

### 4.2. Jump abundance models

The envelope models for AFGL 2591 (Stäuber et al. 2005) suggest a jump in abundance for CO$^+$, SO$^+$ and other molecules at $T = 100$ K where water and H$_2$S evaporate into the gas-phase. The abundances can be several orders of magnitude higher in the inner hot region of the envelope due to the influence of the enhanced water and sulphur abundances. The CN, NO, CO$^+$ and SO$^+$ abundances towards AFGL 2591, W3 IRS5 and IRAS 16293–2422 have therefore been calculated in a first jump model assuming the observed molecular emission to originate mainly from the region with $T \gtrsim 100$ K. The outer abundances $x_{out}$ are assumed to be $10^{-8}$ for NO, $10^{-15}$ for CO$^+$ and $10^{-13}$ for SO$^+$. For CN we assume $x_{out} = 10^{-9}$ for the high-mass sources and $x_{out} = 10^{-11}$ for IRAS 16293–2422.

Abundance enhancements due to the influence of X-rays are expected in the envelope within the inner $\approx 1000$ AU from the central source for $L_X \approx 10^{30}$–$10^{31}$ erg s$^{-1}$ (Stäuber et al. 2005). Since the attenuation of X-rays is dominated by geometric dilution rather than absorption (Stäuber et al. 2006), the X-ray dominated region will be of similar size for all sources. We therefore calculate the abundances in a second jump model assuming the jump to be at 1000 AU. The influence of a strong inner FUV field is restricted to $A_V \lesssim 10$. This is already achieved within a few hundred AU in the high-mass envelopes and a few AU in the low-mass objects. Although high-$J$ lines are sensitive to even such a small region, the predicted line fluxes are too small to account for the observed ones (see also Stäuber et al. 2004).

The inner abundances $x_{100K}$ and $x_{XDR}$ for the models assuming a jump at $T = 100$ K and $r = 1000$ AU, respectively, are presented in Tables 7 and 8. Also listed in the tables are the average hydrogen density and gas temperature in the regions of interest. Some inner abundances in the jump models are more than two orders of magnitude higher compared to the constant abundances in Table 6. However, the jump and constant abundance models are statistically not distinguishable since they produce similar $\chi^2$ values so that all scenarios remain plausible.

## 5. Comparison with chemical models

### 5.1. Abundances and column densities

In comparison with chemical models, both absolute abundances and column densities are relevant. For example, com-parison of the observed values with those of our and other dense PDR models (e.g., Sternberg & Dalgarno 1995) shows good agreement for the abundances of selected species in narrow PDR zones, but they generally underpredict the column densities. Thus, using only the local abundances as a diagnostic of a physical process is not sufficient.

**Table 7.** Inferred inner abundances $x_{100K}$ from jump models assuming the emission to come from the region with $T \gtrsim 100$ K. The average temperature $T_{av}$, hydrogen density $n_{H,av}$ and size (radius) of the inner 100 K region are also given.

| Source | $T_{av}$ (K) | $n_{H,av}$ (cm$^{-3}$) | Radius (AU) | CN | NO | CO$^+$ | SO$^+$ |
|---|---|---|---|---|---|---|---|
| AFGL 2591 | 170 | $4 \times 10^6$ | 1100 | $6 \times 10^{-07}$ | $3 \times 10^{-06}$ | $1 \times 10^{-09}$ | $2 \times 10^{-09}$ |
| W3 IRS 5 | 210 | $1 \times 10^7$ | 5400 | $1 \times 10^{-08}$ | $2 \times 10^{-07}$ | $7 \times 10^{-11}$ | $9 \times 10^{-10}$ |
| IRAS 16293 | 160 | $1 \times 10^9$ | 100 | $3 \times 10^{-09}$ | ... | $6 \times 10^{-11}$ | $\lesssim 1 \times 10^{-10}$ |

**Table 8.** Inferred inner abundances $x_{XDR}$ from jump models assuming the emission to come from the X-ray dominated inner 1000 AU envelope. The average temperature $T_{av}$ and hydrogen density $n_{H,av}$ of the inner 1000 AU region are also given.

| Source | $T_{av}$ (K) | $n_{H,av}$ (cm$^{-3}$) | CN | NO | CO$^+$ | SO$^+$ |
|---|---|---|---|---|---|---|
| AFGL 2591 | 180 | $4 \times 10^6$ | $8 \times 10^{-07}$ | $4 \times 10^{-06}$ | $2 \times 10^{-09}$ | $3 \times 10^{-09}$ |
| W3 IRS 5 | 490 | $6 \times 10^7$ | $5 \times 10^{-07}$ | $5 \times 10^{-06}$ | $5 \times 10^{-10}$ | $3 \times 10^{-08}$ |
| IRAS 16293 | 50 | $2 \times 10^8$ | $8 \times 10^{-11}$ | ... | $2 \times 10^{-12}$ | $\lesssim 5 \times 10^{-12}$ |

The best test of various physical processes is formed by taking a physical model of a source constrained by observational data, and couple it with a chemistry code. AFGL 2591 is the only source in our sample for which an envelope model containing an inner source of FUV or X-ray emission has been combined with chemistry (Stäuber et al. 2004, 2005). Models with only a central FUV source (Stäuber et al. 2004) show that the enhanced region is generally too small to account for the observed column densities (see also Appendix B), unless the photons can escape through cavities and affect a larger column.

Our observed CN and SO$^+$ abundances for AFGL 2591 are comparable to those predicted by the X-ray models of Stäuber et al. (2005) for $L_X \gtrsim 10^{32}$ erg s$^{-1}$. The observed NO abundance is higher than in the models, suggesting an additional NO destruction mechanism not taken into account in the current models. The observed CO$^+$ abundance is several orders of magnitude higher than predicted, however, and is not consistent with the chemical X-ray models.

To compare observations and models for the other sources, a generic grid of CN, NO, SO$^+$ and CO$^+$ abundances has been computed as functions of the X-ray and FUV flux, the gas temperature and the hydrogen density (Appendix B). A summary of the X-ray and/or FUV fluxes consistent with the observational data is given in Table B.2.

It is seen that X-ray models can explain most observed constant fractional abundances for X-ray fluxes $F_X \approx 10^{-4}$–$1$ erg s$^{-1}$ in both low and high-mass sources. The XDR jump abundances in the high-mass objects, however, can only be modeled with $F_X \gtrsim 1$ erg s$^{-1}$. At 1000 AU, such X-ray fluxes correspond to $L_X \gtrsim 10^{33}$ erg s$^{-1}$. This is more than an order of magnitude higher than typical X-ray luminosities observed towards young massive objects (e.g., Townsley 2006). Unless the X-ray luminosity is higher in deeply embedded sources, X-rays are not a plausible explanation for the observed molecular line emission in the high-mass sources. The CN abundances for IRAS 16293–2422 require $F_X \approx 10^{-4} - 10^{-2}$ erg s$^{-1}$, corresponding to $L_X \gtrsim 10^{29}$ erg s$^{-1}$, which is in the range of observed luminosities. FUV models could also explain the CN abundances with a moderate flux ($G_0 \gtrsim 5$) – independent of the gas temperature – but the observed NO, SO$^+$ and CO$^+$ abundances require gas temperatures above $\approx 300$ K if explained only by enhanced UV. The gas temperature around IRAS 16293–2422 is unlikely to be so high (e.g., Schöier et al. 2002), suggesting that X-rays are more important than FUV for IRAS 16293–2422.

### 5.2. Abundance ratios

Rather than trying to match absolute abundances and column densities, one can also compare abundance or column density ratios of two species with chemical models. The advantage is that absolute uncertainties in, for example, rate coefficients common to both molecules or in overall geometry drop out. The disadvantage is that this approach assumes that the two molecules are spatially coexistent, which is not always the case, especially in a layered PDR structure.

In the following sections we discuss constant molecular abundance ratios and study their dependence on X-rays and FUV fields. The CN to HCN ratio, for example, is well known to be a good tracer for enhanced FUV fluxes. In the vicinity of strong FUV fields the CN/HCN ratio is observed to be $\gtrsim 1$ (e.g., Fuente et al. 1993) in good agreement with chemical PDR models (e.g., Jansen et al. 1995; Sternberg & Dalgarno 1995). The observed CN/HCN, CN/NO, CO$^+$/HCO$^+$ and SO$^+$/SO abundance ratios for all sources are presented in Table 9. The abundances of species studied in this paper were taken from radiative transfer models assuming constant fractional abundances as described in Sect. 4.1 (Table 6). Those for HCN, HCO$^+$ and SO are from the literature and are presented in Table 10. They are based on similar radiative transfer models, except for $HCO^+$ and HCN towards W3 IRS5 which were derived from statistical equilibrium calculations at a single temperature and density using an escape probability formalism (Helmich & van Dishoeck 1997).

### 5.2.1. CN to HCN ratios

Our CN/HCN ratios are $\gtrsim 1$ for both high-mass sources and for the Class I object L1489. The ratio is $\approx 1$ for L483 and L723. All other sources have $x(CN/HCN) \lesssim 0.2$ (Table 9).

The CN/HCN ratio is studied as a function of the X-ray flux, the gas temperature and the hydrogen density (Fig. 9; see Appendix B for details on the model). Ratios of the order $\approx 0.1$ are reached for gas temperatures $\lesssim 200$ K, $F_X = 10^{-3}$–$10^{-2}$ erg s$^{-1}$ cm$^{-2}$ and $n_H = 10^6$–$10^7$ cm$^{-3}$. At 1000 AU, this corresponds to reasonable luminosities of $L_X \approx 10^{30}$ erg s$^{-1}$. However, a factor of 1000 higher X-ray fluxes ($F_X \gtrsim 1$ erg s$^{-1}$ cm$^{-2}$) and gas temperatures $T \lesssim 200$ K are needed to give CN/HCN ratios of $\approx 1$ for densities $n_H \gtrsim 10^7$ cm$^{-3}$. X-ray fluxes of the order 1 erg s$^{-1}$ correspond to $L_X \approx 10^{33}$ erg s$^{-1}$ for an XDR region of the size 1000 AU, higher than observed. Lowering the density would lower the X-ray flux required to give a CN/HCN ratio $\gtrsim 1$, but such densities would be below the critical density of the observed lines. Thus, CN/HCN ratios $\gtrsim 1$ are not likely to be due to the influence of X-rays.

The FUV model results are presented in Fig. 10. The CN/HCN ratio is between 1–100 for relatively low FUV fields ($G_0 \lesssim 100$) and temperatures below $\approx 200$ K. The ratio decreases with increasing gas temperature. The observed ratios of $\lesssim 0.2$ require either gas temperatures $T \gtrsim 300$ K for $n_H = 10^6$ cm$^{-3}$ or $G_0 < 5$. For $n_H = 10^7$ cm$^{-3}$, the gas temperature can be lower for FUV fields $G_0 \lesssim 10$.

In summary, low CN/HCN ratios of $\lesssim 0.2$ are consistent with X rays or high-temperature PDRs, whereas high CN/HCN ratios of $\gtrsim 1$ can only be explained with low-temperature PDRs.

### 5.2.2. Other ratios

Our observed $CO^+/HCO^+$ ratios are of the order $10^{-3}$ in the low-mass objects and $\approx 10^{-2}$–$10^{-1}$ in the high-mass sources (Table 9). The X-ray models of Stäuber et al. (2005) predict $CO^+/HCO^+$ ratios of $\lesssim 10^{-6}$. The FUV envelope models of Stäuber et al. (2004) show ratios between $10^{-5}$–$10^{-2}$ for $G_0 = 10$–$10^5$. Our observed values are in the range of our FUV models.

The $SO^+/SO$ ratio for the high-mass objects is 6–7 × $10^{-3}$. The upper limit for the low-mass YSOs is $\approx 10$ times less. The X-ray envelope models of AFGL 2591 have $SO^+/SO \approx 10^{-3}$, comparable to the observations. Our PDR models predict ratios between $\approx 1$–$10^3$ in the innermost region, for $G_0 = 10$–$10^4$. If the observed $SO^+$ abundance were from a PDR, the SO emission would have to come from a region that is not affected by the strong FUV radiation. Assuming that they originate from the same region, only X-rays can explain the observations.

**Table 9.** Observed abundance ratios assuming constant abundances.

| Source | CN/HCN | CN/NO | $CO^+/HCO^+$ | $SO^+/SO$ |
|---|---|---|---|---|
| AFGL 2591 | 1.15 | 0.74 | 0.016 | 0.006 |
| W3 IRS 5 | 1.80 | 0.21 | 0.066 | 0.007 |
| IRAS 16293 | 0.05 | ... | 0.001 | < 0.001 |
| N1333-I2 | 0.20 | 0.16 | 0.002 | < 0.001 |
| N1333-I4A | 0.12 | 0.03 | < 0.002 | < 0.001 |
| L1448-C | 0.20 | ... | < 0.002 | < 0.007 |
| L483 | 0.75 | ... | < 0.004 | ... |
| L723 | 0.92 | ... | < 0.005 | < 0.017 |
| SMM4 | 0.08 | ... | < 0.001 | ... |
| L1489 | 3.85 | ... | ... | ... |

**Table 10.** Observed fractional abundances from literature.

| Source | HCN ×$10^{-9}$ | $HCO^+$ ×$10^{-9}$ | SO ×$10^{-9}$ |
|---|---|---|---|
| AFGL 2591 | 20.0[a] | 10.0[a] | 10.0[b] |
| W3 IRS 5 | 2.0[c] | 0.73[c] | 10.0[b] |
| IRAS 16293 | 1.1[d] | 1.4[d] | 4.4[d] |
| N1333-I2 | 2.0[e] | 3.3[e] | 3.4[e] |
| N1333-I4A | 0.36[e] | 0.43[e] | 4.6[e] |
| L1448-C | 5.4[e] | 9.1[e] | 1.4[e] |
| L483 | 2.0[e] | 2.0[e] | 0.29[e] |
| L723 | 1.0[e] | 4.1[e] | 2.4[e] |
| SMM4 | 2.0[f] | 1.0[f] | ... |
| L1489 | 0.65[e] | 18.0[e] | 2.0[e] |

[a] van der Tak et al. (2000); [b] van der Tak et al. (2003) [c] Helmich et al. (1997); [d] Schöier et al. (2002); [e] Jørgensen et al. (2004); [f] Hogerheijde et al. (1999)

## 6. Discussion

Comparison of observations with chemical models for high mass sources indicates that the observed CN, $CO^+$ and $SO^+$ abundances, as well as the CN/HCN and $CO^+/HCO^+$ ratios, are best explained by enhanced FUV fluxes in a PDR model with temperatures of a few hundred K. Only the $SO^+/SO$ ratio is inconsistent with this conclusion, suggesting that the bulk of the $SO^+$ and SO are not co-located. The observed absolute column densities indicate that such a PDR must be extended on scales of at least 1000 AU, i.e., much more than expected from a homogeneous spherically symmetric envelope. Possible geometries will be discussed in Sect. 6.1. The required X-ray fluxes to reproduce the abundances and ratios are at least an order of magnitude higher than observed toward young massive objects, making this scenario less likely for the species observed in this work.

For low-mass objects, X-rays with luminosities $L_X \gtrsim 10^{29}$ erg s$^{-1}$ are the likely explanation of the observed CN, $CO^+$ and $SO^+$ abundances, since FUV models require unrealistically high temperatures of >300 K over large volumes, except for the case of CN. Only sources with CN/HCN ratios ≳ 1 and/or detected CO$^+$ may require an additional PDR contribution.

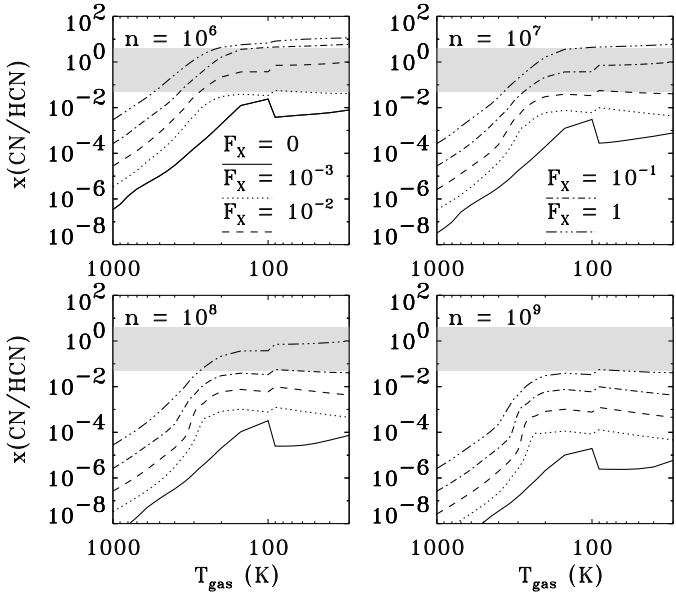

**Fig. 9.** Modeled CN/HCN ratio as functions of the X-ray flux (erg s$^{-1}$ cm$^{-2}$), gas temperature and total hydrogen density (cm$^{-3}$). The shaded region indicates the observed ratios (Table 9).

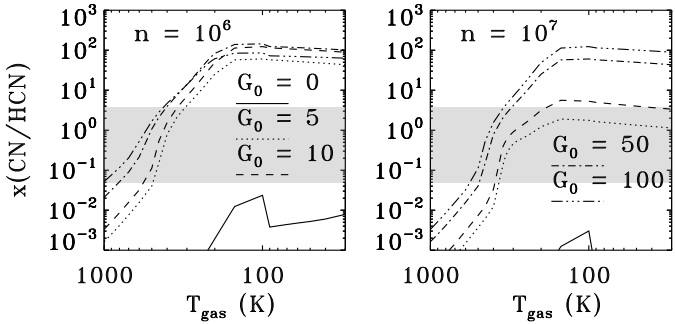

**Fig. 10.** Modeled CN/HCN ratio as functions of the FUV field strength, gas temperature and total hydrogen density (cm$^{-3}$). The shaded region indicates the observed ratios (Table 9).

NO is consistently overproduced in the models, suggesting that some destruction or depletion mechanism is missing in the networks.

### 6.1. Geometric effects

FUV fields from the protostar can only explain our observations if there is a low-opacity region in the envelope that allows the FUV photons to escape and affect gas on larger scales. One possibility is the scattering of FUV photons in outflow cavities as suggested by Spaans et al. (1995). Another option is a geometry as presented in Fig. 11, where the FUV photons could also impact the envelope directly on large scales without being scattered. In this scenario, the protostar is surrounded by a (flaring) disk and low density outflows. In low-mass objects, such a geometry is expected to be shaped by disk-driven winds (e.g., Shu et al. 1994). A region in the envelope with low optical depth was suggested for AFGL 2591 by van der Tak et al. (1999). Images from the Spitzer Space Telescope of high-mass objects generally show a patchy structure with regions through which FUV photons can escape (e.g., Churchwell et al. 2004, 2006).

**Table 11.** Summary of processes predicted to be involved in the formation of each molecule.

| Molecule | high mass YSOs | low-mass YSOs |
|---|---|---|
| CN | FUV cavity walls | X-rays |
| | X-rays | |
| NO | X-rays | X-rays |
| | FUV cavity walls | |
| SO$^+$ | FUV cavity walls | X-rays |
| | X-rays | |
| CO$^+$ | FUV cavity walls | inner region X-rays |
| | | inner region FUV |

The same scenario could be possible for X-rays. X-rays, however, do not depend on geometry as much as FUV fields due to the smaller absorbing cross sections. An X-ray absorbing column density of $N(H_2) \lesssim 10^{23}$ cm$^{-2}$ reduces the X-ray flux only by a factor of a few but not by orders of magnitude (Stäuber et al. 2005, 2006). The main reducing parameter for the X-ray flux is the geometric dilution ($F_X \propto r^{-2}$). The X-ray enhanced cavity walls are thus of minor importance compared to the quiescent bulk envelope irradiated by X-rays.

Alternatively, X-rays may be powered by fast winds or jets from the embedded protostar. In this scenario, the source of X-ray emission are shocks shifted from the central position which may even be multiple in nature. X-rays can then impact the envelope on large scales with luminosities between $L_X \approx 10^{29} - 10^{33}$ erg s$^{-1}$ (e.g., Favata et al. 2002; Ezoe et al. 2006). The thermal X-ray spectrum is given by the temperature of the shocked gas ($\approx 1$ keV $(V/1000\,\text{km s}^{-1})^2$). An ionized wind with $V \approx 500$ km s$^{-1}$ has been observed in the [SII] $\lambda$6731 line towards AFGL 2591 by Poetzel et al. (1992). Such a velocity would lead to the emission of soft X-rays ($\approx 3 \times 10^6$ K) on large scales if it came to a shock with the surrounding material. However, the investigation of this scenario is beyond the scope of this paper.

The possible scenarios are further discussed separately for the high and low-mass objects in the next two sections. A summary of the processes that we believe are traced by each molecule for the low and high-mass objects is presented in Table 11.

### 6.2. High-mass objects

As argued above, FUV fields with $G_0 \gtrsim 5$ and high gas temperatures ($T \gtrsim 300$ K) are the most likely explanation for the observations, consistent with the outflow cavity scenario. The question is whether a source like AFGL 2591 can maintain high FUV fields and gas temperatures out to such large distances from the protostar. To estimate the FUV flux for AFGL 2591 at $r_{\text{beam}} = $ HPBW/2 $\approx 7000$ AU, we assume a black body spectrum with a stellar temperature $T = 3 \times 10^4$ K and $L_{bol} = 2 \times 10^4 L_\odot$ (van der Tak et al. 1999). This corresponds to $G_0(r_{beam}) \approx 2 \times 10^5$ ($G_0 = 1$ corresponds to $1.6 \times 10^{-3}$ erg s$^{-1}$ cm$^{-2}$, Habing 1968). Material in the outflows and geometric dilution due to a non-perpendicular impact angle will reduce this FUV flux by a factor of $\approx 10$–$1000$, depending mainly on the outflow density. However, the effective FUV flux at $r_{beam}$ can still be between $G_{0,eff} \approx 10$–$10^4$. If the heating of the gas along the outflow walls is provided by FUV, a temperature of $\approx 300$ K requires $G_0 \approx 10^4$ (e.g., Sternberg & Dalgarno 1995). This implies $A_V \lesssim 2$ at $r_{beam}$ and the material in the outflow is therefore suggested to have a density of $\lesssim 4 \times 10^4$ cm$^{-3}$.

To further test whether the FUV scattering scenario of Spaans et al. (1995) is possible for the high-mass objects, the Spaans et al. model is scaled to the luminosity and radial size of AFGL 2591. A 3D model of the region is constructed with a spherical envelope and an outflow cavity, assuming a 30 degree opening angle. This leads to a region of high FUV fields ($G_0 \approx 1000$) and $T \gtrsim 100$ K out to $r_{beam}$ along the outflow walls. By convolving this envelope model with the JCMT beam for different angles of telescope pointing relative to the direction of outflow, it is seen that $6-43\%$ of the volume is in the high FUV region, assuming the molecular line emission to be optically thin. Applying this result to the fractional abundances from the FUV model results of Stäuber et al. (2004) leads to $x(CN) \approx 6 \times 10^{-9} - 4 \times 10^{-8}$, $x(NO) \approx 6 \times 10^{-9} - 4 \times 10^{-8}$, $x(SO^+) \approx 6 \times 10^{-11} - 4 \times 10^{-10}$, $x(CO^+) \approx 6 \times 10^{-11} - 4 \times 10^{-10}$. These values are in good agreement to those observed towards AFGL 2591. The observed column densities can easily be achieved if we assume the density profiles in the outflow walls to be the same as that in the envelope ($n \propto r^{-1}$; van der Tak et al. 1999). If the species trace an FUV enhanced region at few $A_V$ away from the outflow cavity wall, the center velocity can be expected to be the one of the system. In addition, such a relatively thin layer of gas would lead to the observed narrow lines (Sect. 3.1). This simple model shows that strong FUV fields from high-mass objects can explain the observed features if the FUV photons travel through the outflow cavities and affect the envelope at large distances either due to scattering or direct impact.

If the species observed in this work indeed trace FUV enhanced regions along the outflow walls, the question is then whether or not X-rays are important for the overall chemistry in envelopes around high-mass protostars. In the case of our sample of species, approximately 10–20% of the emission could be attributed to the influence of X-rays. This is not likely to be the case for pure X-ray tracers. Stäuber et al. (2005) found that a large number of other molecules were better fitted in models assuming a central X-ray source. Only a few species among them, however, were tracers of the FUV field. Species like $N_2H^+$ or $HCO^+$, for example, are reduced in FUV enhanced regions (eg., Jansen et al. 1994) but were well fitted within the X-ray models. Thus, X-rays are still necessary to properly understand the envelope chemistry.

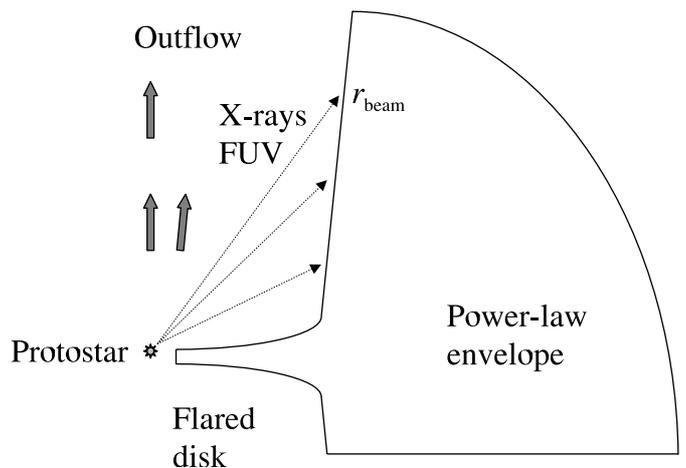

**Fig. 11.** Schematic drawing of a possible geometry for the impact of X-rays and FUV radiation on the envelope surrounding a low-mass protostar with a flaring disk. The dotted arrows indicate the X-rays and FUV photons. The plot is not to scale.

### 6.3. Low-mass objects

#### 6.3.1. FUV

Low-mass objects emit much less UV photons than high-mass sources due to their lower surface temperatures. However, accretion may heat the infalling ionized gas close to the star to to high temperatures and increase the FUV flux. Taking this into account, the estimated FUV flux is $G_0 \lesssim 10^4$ at 100 AU for a source with $L_{bol} \approx 20 L_\odot$ and an effective temperature $T_{eff} \approx 7000$ K. Bergin et al. (2003) argued that $G_0$ may be only a few hundred, albeit at later stages, so that the estimated value above can be regarded as an upper limit for $G_0$ at 100 AU. Considering absorption in the outflows and geometric dilution, the effective FUV field at $r_{beam}$ is $G_{0,eff} \lesssim$ a few. Also, to explain the observed CO$^+$ (and SO$^+$) abundance with enhanced FUV fluxes, the gas temperature would have to be $\gtrsim 300$ K. Such high temperatures are not predicted by any models. The CO$^+$ emission in IRAS 16293–2422 is thus not likely to be due to the influence of a central FUV field.

CN appears to be a different case. Observations towards the Class 0 object L483 by Jørgensen (2004) indicate an extended CN emission morphology which is suggestive of FUV enhancement along the cavity walls. It has been shown that low FUV fluxes can already enhance CN by more than an order of magnitude (Fig. B.2). However, the FUV radiation field of low-mass YSOs may not have enough UV photons at wavelengths $< 1100$ Å required to photodissociate CN so that its abundance may be even higher than predicted by our models. On the other hand, outflows in young low-mass objects appear to be more collimated than in high-mass objects, so that the outflow cavity walls may contribute only a few percent to the observed emission in a 14" beam, depending on the line of sight. The bulk CN emission would thus still have to come from the X-ray enhanced envelope in some sources. High spatial resolution observations of high-excitation CN lines are needed to further investigate this question.

CO$^+$ and CN could also trace an FUV enhanced gas in the protostellar hole region, in the inner edge of the envelope or in the disk ($r \lesssim 500$ AU). The temperature requirement of $T \gtrsim 300$ K for CO$^+$, however, is not consistent with the dust radiative transfer models for the low-mass sources (Jørgensen et al. 2002; Schöier et al. 2002). To provide the additional heating through FUV photons at $n_H = 10^6$–$10^7$ cm$^{-3}$ (comparable to upper disk layers), requires $G_0 \gtrsim 10^4$ (e.g., Sternberg & Dalgarno 1995), at the upper limit of what is likely.

### 6.3.2. X-rays

Given the problems with the FUV models, the observed emission of CN in all low-mass objects and CO$^+$ in IRAS 16293–2422 may reflect the influence of X-rays in upper disk layers or in the innermost part of the envelope.

Assuming CN to originate from the envelope, the protostellar X-ray luminosity can be estimated from the model results (Sect. 5) and the observed abundances (Table 6). For simplicity, $n_H$ is taken to be $\approx n(r_{beam})$ and the gas temperature is assumed to be $T \lesssim 100$ K. Table 12 shows the results of this comparison for CN. Since X-rays will be attenuated by the gas between the protostar and $r_{beam}$, the X-ray luminosities in Table 12 are given as lower limits. To explain the derived constant fractional CO$^+$ abundance for IRAS 16293–2422 in the same way, the X-ray luminosity would have to be at least two orders of magnitude higher.

If we assume the observed emission from IRAS 16293–2422 to come from upper disk layers or gas in the protostellar hole region ($r \lesssim 500$ AU) with $n_H \approx 10^6$ cm$^{-3}$, the X-ray luminosity would have to be $L_X \approx 10^{30}$–$10^{32}$ erg s$^{-1}$ to account for the observed CN and CO$^+$ emission.

We conclude that CN and CO$^+$ are tracing an X-ray enhanced region close to the protostar since FUV enhanced outflow walls may contribute only a few percent to the observed emission and since CO$^+$ requires high gas temperatures in the FUV scenario. The estimated luminosities in all X-ray scenarios are between a few $\times 10^{29}$ – $10^{32}$ erg s$^{-1}$ which is in good agreement to observations of Class I or older protostars. Class 0 objects may therefore emit X-rays at similar levels as later type YSOs.

## 7. Conclusion

Several molecular ions and radicals that are thought to trace FUV and/or X-rays have been observed and detected towards both low and high-mass star-forming regions. For the FU Orionis object V1057 Cyg, continuum SCUBA observations have been carried out and the results are used to constrain the one dimensional physical structure of the envelope. The fractional molecular abundances are estimated through Monte Carlo line radiative transfer modeling. Chemical models are presented and compared to the observed abundances.

The observed CN, SO$^+$ and CO$^+$ abundances and column densities in the high-mass objects are best explained by FUV enhanced outflow cavity walls with $G_0 \gtrsim 10$ and $T \gtrsim 300$ K. Low-mass objects are less FUV active and have generally more collimated outflows. In these sources, the observed features are best attributed to the influence of X-rays, although FUV fields may contribute for sources with high CN/HCN ratios. The observed abundances imply X-ray fluxes for Class 0 objects at similar levels as more evolved low-mass protostars.

**Table 12.** Estimates of the X-ray luminosity in low-mass objects from the observed CN abundances.

| Source | $r_{beam}$ AU | $n_H(r_{beam})$ cm$^{-3}$ | X-ray luminosity $L_X$ erg s$^{-1}$ |
|---|---|---|---|
| IRAS 16293 | 1120 | 10$^7$ | $\gtrsim 4 \times 10^{29}$ |
| N1333-I2 | 1540 | 10$^6$ | $\gtrsim 7 \times 10^{29}$ |
| N1333-I4A | 1540 | 10$^7$ | $\gtrsim 7 \times 10^{29}$ |
| L1448-C | 1540 | 10$^6$ | $\gtrsim 7 \times 10^{30}$ |
| L483 | 1400 | 10$^6$ | $\gtrsim 6 \times 10^{30}$ |
| L723 | 2100 | 10$^6$ | $\gtrsim 1 \times 10^{31}$ |
| SMM4 | 1750 | 10$^6$ | $\gtrsim 1 \times 10^{30}$ |
| V1057 Cyg | 4550 | 10$^5$ | $\gtrsim 6 \times 10^{30}$ |
| L1489 | 980 | 10$^5$ | $\gtrsim 8 \times 10^{29}$ |

Observation of other molecules in the same sources may require a mix of FUV and X ray processes. Which of the two dominates for a particular species cannot be clearly determined from unresolved single-dish data. Future interferometric observations with e-SMA or ALMA should clarify this question. Also, chemical models that take the 3D geometry of the source into account are needed.

*Acknowledgements.* The authors are grateful to the JCMT staff, in particular Remo Tilanus, for excellent support and assistance. We also thank John Black for providing the molecular CO$^+$ data. This work was partially supported under grants from The Research Corporation (SDD). The research of JKJ was supported by NASA Origins Grant NAG5-13050. Astrochemistry in Leiden is supported by the Netherlands Research School for Astronomy (NOVA) and by a Spinoza grant from the Netherlands Organization for Scientific Research (NWO).

## References

Ábrahám, P., Kóspál, Á., Csizmadia, S., Kun, M., Moór, A., & Prusti, T. 2004, A&A, 428, 89
Bergin, E., Calvet, N., D'Alessio, P., & Herczeg, G. J. 2003, ApJ, 591, L159
Black, J. H. 1998, Chemistry and Physics of Molecules and Grains in Space. Faraday Discussions No. 109, 257
Blake, G. A., Sutton, E. C., Masson, C. R., & Phillips, T. G. 1987, ApJ, 315, 621
Boonman, A. M. S., Doty, S. D., van Dishoeck, E. F., Bergin, E. A., Melnick, G. J., Wright, C. M., & Stark, R. 2003a, A&A, 406, 937
Boonman, A. M. S., van Dishoeck, E. F., Lahuis, F., & Doty, S. D. 2003b, A&A, 399, 1063
Boonman, A. M. S. & van Dishoeck, E. F. 2003, A&A, 403, 1003
Ceccarelli, C., Hollenbach, D. J., & Tielens, A. G. G. M. 1996, ApJ, 471, 400
Ceccarelli, C., Caux, E., White, G. J., et al. 1998, A&A, 331, 372
Charnley, S. B. 1997, ApJ, 481, 396
Churchwell, E., et al., 2004, ApJS, 154, 322
Doty, S. D., van Dishoeck, E. F., van der Tak, F. F. S., & Boonman, A. M. S. 2002, A&A, 389, 446

Doty, S. D., Schöier, F. L., & van Dishoeck, E. F. 2004, A&A, 418, 1021
Ezoe, Y., Kokubun, M., Makishima, K., Sekimoto, Y., & Matsuzaki, K. 2006, ApJ, 638, 860
Favata, F., Fridlund, C. V. M., Micela, G., Sciortino, S., & Kaas, A. A. 2002, A&A, 386, 204
Forbrich, J., Preibisch, T., & Menten, K. M. 2006, A&A, 446, 155
Fuente, A., Martin-Pintado, J., Cernicharo, J., & Bachiller, R. 1993, A&A, 276, 473
Fuente, A., Rodrıguez-Franco, A., Garcıa-Burillo, S., Martın-Pintado, J., & Black, J. H. 2003, A&A, 406, 899
Fuente, A., García-Burillo, S., Gerin, M., Rizzo, J. R., Usero, A., Teyssier, D., Roueff, E., & Le Bourlot, J. 2006, A&A, 641, L105
Green, S., & Chapman, S. 1978, ApJS, 37, 169
Habing, H. J. 1968, Bull. Astron. Inst. Netherlands, 19, 421
Hamaguchi, K., Corcoran, M. F., Petre, R., White, N. E., Stelzer, B., Nedachi, K., Kobayashi, N., & Tokunaga, A. T. 2005, ApJ, 623, 291
Helmich, F. P., & van Dishoeck, E. F. 1997, A&AS, 124, 205
Hofner, P., Delgado, H., Whitney, B., Churchwell, E., & Linz, H. 2002, ApJ, 579, L95
Hogerheijde, M. R., van Dishoeck, E. F., Salverda, J. M., & Blake, G. A. 1999, ApJ, 513, 350
Hogerheijde, M. R., & van der Tak, F. F. S. 2000, A&A, 362, 697
Hogerheijde, M. R. 2001, ApJ, 553, 618
Imanishi, K., Koyama, K., & Tsuboi, Y. 2001, ApJ, 557, 747
Jansen, D. J., van Dishoeck, E. F., & Black, J. H. 1994, A&A, 282, 605
Jansen, D. J., van Dishoeck, E. F., Black, J. H., Spaans, M., & Sosin, C. 1995, A&A, 302, 223
Jørgensen, J. K., Schöier, F. L., & van Dishoeck, E. F. 2002, A&A, 389, 908
Jørgensen, J. K. 2004, A&A, 424, 589
Jørgensen, J. K., Schöier, F. L., & van Dishoeck, E. F. 2004, A&A, 416, 603
Kenyon, S. J., & Hartmann, L. W. 1991, ApJ, 383, 664
Latter, W. B., Walker, C. K., & Maloney, P. R. 1993, ApJ, 419, L97
Liszt, H. S., & Turner, B. E. 1978, ApJ, 224, L73
Lorenzetti, D., Giannini, T., Nisini, B., et al. 2000, A&A, 357, 1035
Millar, T. J., Adams, N. G., Smith, D., Lindinger, W., & Villinger, H. 1986, MNRAS, 221, 673
Poetzel, R., Mundt, R., & Ray, T. P. 1992, A&A, 262, 229
Pontoppidan, K. M., van Dishoeck, E. F., & Dartois, E. 2004, A&A, 426, 925
Preibisch, T., Kim, Y. C., Favata, F., et al. 2005, ApJS, 160, 401
Sandell, G., & Weintraub, D. A. 2001, ApJS, 134, 115
Savage, C., Apponi, A. J., & Ziurys, L. M. 2004, ApJ, 608, L73
Savage, C., & Ziurys, L. M. 2004, ApJ, 616, 966
Schöier, F. L., Jørgensen, J. K., van Dishoeck, E. F., & Blake, G. A. 2002, A&A, 390, 1001
Shu, F., Najita, J., Ostriker, E., Wilkin, F., Ruden, S., & Lizano, S. 1994, ApJ, 429, 781
Spaans, M., Hogerheijde, M. R., Mundy, L. G., & van Dishoeck, E. F. 1995, ApJ, 455, L167
Stäuber, P., Doty, S. D., van Dishoeck, E. F., Jørgensen, J. K., & Benz, A. O. 2004, A&A, 425, 577
Stäuber, P., Doty, S. D., van Dishoeck, E. F., & Benz, A. O. 2005, A&A, 440, 949
Stäuber, P., Jørgensen, J. K., van Dishoeck, E. F., Doty, S. D., & Benz, A. O. 2006, A&A, 453, 555
Sternberg, A., & Dalgarno, A. 1995, ApJS, 99, 565
Townsley, L., 2006, in the STScI May Symposium, "Massive Stars: From Pop III and GRBs to the Milky Way", ed. M. Livio, (astro-ph/0608173)
Turner, B. E. 1992, ApJ, 396, L107
Turner, B. E. 1994, ApJ, 430, 727
van der Tak, F. F. S., van Dishoeck, E. F., Evans, N. J., Bakker, E. J., & Blake, G. A. 1999, ApJ, 522, 991
van der Tak, F. F. S., van Dishoeck, E. F., Evans, N. J., & Blake, G. A. 2000, ApJ, 537, 283
van der Tak, F. F. S., Boonman, A. M. S., Braakman, R., & van Dishoeck, E. F. 2003, A&A, 412, 133
van Dishoeck, E. F., Blake, G. A., Jansen, D. J., & Groesbeck, T. D. 1995, ApJ, 447, 760
Ziurys, L. M., McGonagle, D., Minh, Y., & Irvine, W. M. 1991, ApJ, 373, 535

# Online Material

## Appendix A: CO⁺ anomalous excitation

A problem in the interpretation of the observed CO$^+$ emission is the question of how the molecule is excited. CO$^+$ reacts quickly ($\approx 10^{-9}$ cm$^3$ s$^{-1}$) with atomic or molecular hydrogen. Instead of being collisionally excited, CO$^+$ is more likely to be destroyed on virtually every collision with H and H$_2$ (Black 1998). This is not the case for the other observed molecules. The reaction of SH$^+$ with H$_2$, for example, is endothermic with an activation barrier of $\approx 6000$ K (Millar et al. 1986). We assume that this holds also for SO$^+$.

CO$^+$ could be produced in an excited level and the observed emission may be coupled to its formation. If CO$^+$ is not collisionally excited, however, the fractional abundances derived from the radiative transfer modeling do not represent those of CO$^+$. The column densities are therefore calculated using the following expression that is independent of the excitation mechanism and can be obtained by simply integrating the standard radiative transfer equations assuming the line to be optically thin (Blake et al. 1987):

$$N = \frac{1.94 \times 10^3 \nu^2 Q(T_{\rm ex})}{g_{\rm u} A_{\rm ul}} e^{E_{\rm u}/kT_{\rm ex}} \int T_{\rm MB}\, dV, \quad [{\rm cm}^{-2}] \quad ({\rm A.1})$$

where $\nu$ is the line frequency in GHz, $Q(T_{\rm ex})$ is the partition function at the excitation temperature $T_{\rm ex}$, $A_{\rm ul}$ is the Einstein spontaneous emission coefficient in $s^{-1}$, $E_{\rm u}$ and $g_{\rm u}$ are the upper energy level and statistical weight, respectively. The integrated intensity is in K km s$^{-1}$. The results are critically dependent on the excitation temperature, which we do not know. The excitation temperature derived in PDRs by Fuente et al. (2003) is $T_{\rm ex}({\rm CO}^+) = 10$ K. Since our observed transitions are at higher frequencies with $\approx 10$ K higher upper energy levels, the excitation temperature is likely to be higher than that derived by Fuente et al. (2003). We therefore assume $T_{\rm ex}({\rm CO}^+) = 20$ K, which is also consistent with the average excitation temperature derived from the radiative transfer models assuming constant fractional abundances. The results for the CO$^+$($3\,\frac{5}{2}$ - $2\,\frac{3}{2}$) line at 353.741 GHz are presented in Table A.1. The column densities for $T_{\rm ex} = 10$ K are $\approx 3$ times higher and $\approx 30\%$ higher for $T_{\rm ex} = 100$ K. They are $\approx 10$–15% lower for $T_{\rm ex} = 30$–50 K. The fractional abundances are calculated using the integrated hydrogen column densities presented in Table 2. H$_2$ column densities averaged over a 14″ beam may be somewhat lower and the derived fractional abundances can thus be regarded as lower limits. The abundances in Table A.1 are approximately an order of magnitude lower than those inferred from the radiative transport models (Table 6). In case of the low-mass YSO IRAS 16293–2422, the difference is only a factor of four.

## Appendix B: General parameter study

In this section, we compare our observed abundances to those of chemical X-ray and FUV models. As for the case of H$_2$O (Stäuber et al. 2006), abundance results are presented for a general range of parameters. The general parameter study is useful because it covers the entire range of temperatures and densities expected in protostellar environments without assigning them to a specific component like the envelope, disk or dense outflows. Similarities and differences between our models and those previously published are discussed in Stäuber et al. (2004, 2005). In particular, the presence of evaporated H$_2$O at high abundances in the inner envelope results in different chemical characteristics compared with traditional PDR and X-ray dominated region (XDR) models.

The initial chemical abundances for the general parameter study are listed in Table B.1. They are taken to be the same as those in the models for IRAS 16293–2422 by Doty et al. (2004). Many species are taken to be initially frozen out onto grains in the cold part and assumed to evaporate instantaneously into the gas-phase at $T_{\rm ev}$ (Table B.1). However, no subsequent adsorption or desorption of gas-phase species is taken into account. The gas temperature is assumed to be coupled to the dust temperature and independent of the X-ray or FUV flux. The X-ray flux ($F_{\rm X} \propto L_{\rm X} r^{-2}$) is normalized in such a way that a typical protostellar X-ray luminosity of $L_{\rm X} = 10^{31}$ erg s$^{-1}$ corresponds to an X-ray flux of $F_{\rm X} = 1$ erg s$^{-1}$ cm$^{-2}$ at an arbitrary distance of $r \approx 56$ AU from the central star. The temperature for the thermal X-ray spectrum is assumed to be $3 \times 10^7$ K. The X-ray flux is further assumed to be attenuated by a hydrogen column density of $5 \times 10^{21}$ cm$^{-2}$. The FUV flux is varied between $G_0 = 0$–100 for $A_{\rm V} = 0$. A summary of the model results is given in Table B.2.

### B.1. CN

The observed constant fractional abundances are between a few $\times 10^{-11}$–$10^{-8}$ (Table 6) with an average abundance of $10^{-9}$ for the low-mass objects and $10^{-8}$ for the high-mass sources. The jump abundances $x_{\rm 100K}$ are between a few $\times 10^{-9}$–$10^{-7}$ (Table 7) and $x_{\rm XDR}$ ranges from $8 \times 10^{-11}$ to $8 \times 10^{-7}$ (Table 8). Multiplying the constant fractional abundances with the H$_2$ column density (Table 2) for each source leads to CN column densities $N({\rm CN}) \approx 10^{14}$ cm$^{-2}$ for most low-mass sources and $\approx 10^{15}$ cm$^{-2}$ for the high-mass objects and the low-mass YSO L483.

The results of the general parameter study for CN are presented in Fig. B.1 for the X-ray models and in Fig. B.2 for

**Table A.1.** Column densities $N({\rm CO}^+)$ and fractional abundances $x({\rm CO}^+)$ calculated with Eq. (A.1) assuming $T_{\rm ex}({\rm CO}^+) = 20$ K. The upper limits correspond to $3\sigma$ in line flux.

| Source | $N({\rm CO}^+)$ $10^{12}$ cm$^{-2}$ | $x({\rm CO}^+)$ |
|---|---|---|
| AFGL 2591 | 1.0 | $1.0 \times 10^{-11}$ |
| W3 IRS 5 | 1.7 | $2.5 \times 10^{-12}$ |
| IRAS 16293 | 0.6 | $3.9 \times 10^{-13}$ |
| N1333-I2 | 0.2 | $3.3 \times 10^{-13}$ |
| N1333-I4A | < 0.1 | < $6.0 \times 10^{-14}$ |
| L1448-C | < 0.1 | < $6.0 \times 10^{-13}$ |
| L483 | < 0.1 | < $9.0 \times 10^{-14}$ |
| L723 | < 0.1 | < $3.0 \times 10^{-13}$ |
| SMM4 | < 0.1 | < $9.0 \times 10^{-14}$ |
| V1057 Cyg | < 0.1 | < $2.0 \times 10^{-12}$ |

**Table B.1.** Assumed initial abundances and cosmic-ray ionization rate for the chemical models (Figs. B.1–10).

| Species | Initial abundance | Ref. |
|---|---|---|
| Initial abundances $T > 100$ K | | |
| $H_2$ | 1 | |
| CO | $2.0 \times 10^{-4}$ | a |
| $CO_2$ | $3.0 \times 10^{-5}$ | b |
| $H_2O$ | $1.5 \times 10^{-4}$ | c |
| $H_2S$ | $10^{-8}$ | d |
| $H_2CO$ | $8.0 \times 10^{-8}$ | d |
| $N_2$ | $7.0 \times 10^{-5}$ | e |
| $CH_4$ | $10^{-7}$ | e |
| $C_2H_4$ | $8.0 \times 10^{-8}$ | e |
| $C_2H_6$ | $10^{-8}$ | e |
| $CH_3OH$ | $1.5 \times 10^{-7}$ | d |
| O | 0.0 | e |
| S | 0.0 | e |
| Initial abundances $T < 100$ K | | |
| $H_2$ | 1 | |
| CO | $2.0 \times 10^{-4}$ | d |
| $CO_2$ | 0.0 | f |
| $H_2O$ | 0.0 | f |
| $H_2S$ | 0.0 | f |
| $N_2$ | $7.0 \times 10^{-5}$ | e |
| $CH_4$ | $10^{-7}$ | e |
| $C_2H_4$ | $8.0 \times 10^{-8}$ | e |
| $C_2H_6$ | $10^{-8}$ | e |
| $H_2CO$ ($60 < T(K) < 100$) | $8.0 \times 10^{-8}$ | d |
| $H_2CO$ ($T(K) < 60$) | 0.0 | d |
| $CH_3OH$ ($60 < T(K) < 100$) | $1.5 \times 10^{-7}$ | d |
| $CH_3OH$ ($T(K) < 60$) | 0.0 | d |
| O | $1.0 \times 10^{-4}$ | d |
| S | $6.0 \times 10^{-9}$ | g |
| Cosmic-ray ionization rate $\zeta_{cr}$ (s$^{-1}$) | $0.8 \times 10^{-17}$ | h |

All abundances are relative to molecular hydrogen. [a] Jørgensen et al. (2002), [b] Boonman et al. (2003b), [c] Boonman & van Dishoeck (2003), [d] Doty et al. (2004), [e] Charnley (1997), [f] assumed to be frozen-out or absent in cold gas-phase, [g] Doty et al. (2002), [h] Stäuber et al. (2006).

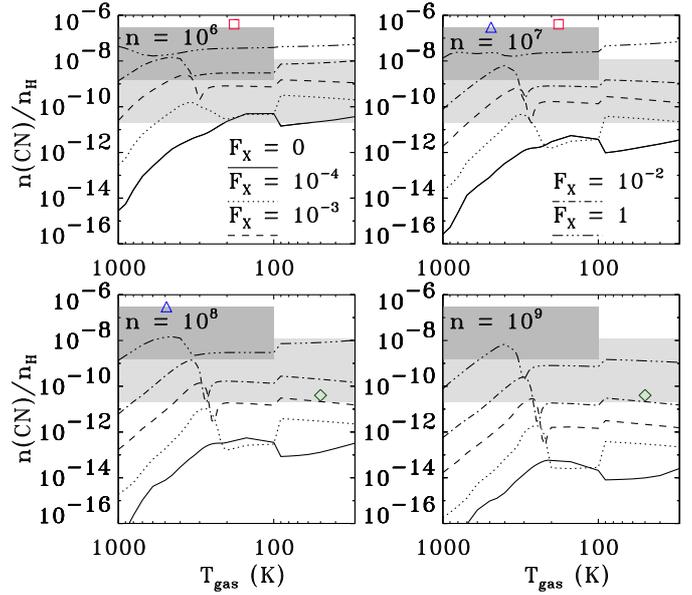

**Fig. B.1.** Modeled CN abundances as a function of the X-ray flux (erg s$^{-1}$ cm$^{-2}$), gas temperature and total hydrogen density (cm$^{-3}$). The light shaded region indicates the observed constant abundances. The dark shaded region indicates the range of $x_{100K}$. The square indicates $x_{XDR}$ for AFGL 2591 at the position of the average temperature and density in this region (Table 8); the triangle represents W3 IRS5 and the diamond represents IRAS 16293–2422.

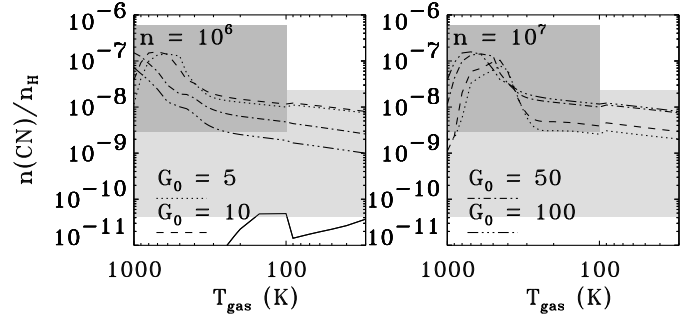

**Fig. B.2.** Modeled CN abundances as a function of the FUV radiation field $G_0$, gas temperature and total hydrogen density (cm$^{-3}$). The light shaded region indicates the observed constant abundances. The dark shaded region indicates the range of $x_{100K}$.

different FUV fields for $t = 5 \times 10^4$ yrs. Compared to models for $t = 10^4$ yrs and $t = 10^5$ yrs, the difference in the CN abundances is minor. In general, the temperature dependence of CN is fairly small in the X-ray models for $T \lesssim 400$ K. At higher temperatures and $F_X < 1$ erg s$^{-1}$ cm$^{-2}$, CN is destroyed by $H_2$, forming HCN. Models without X-rays have $x(CN) < 10^{-10}$ for $n_H = 10^6$–$10^8$ cm$^{-3}$. X-ray fluxes between $10^{-4}$–1 erg s$^{-1}$ cm$^{-2}$ lead to $x(CN) \approx 10^{-10}$–$10^{-7}$ – the approximate range of observed CN abundances. The $x_{XDR}$ jump abundance of IRAS 16293–2422 can be modeled with $F_X \gtrsim 10^{-3}$ erg s$^{-1}$ cm$^{-2}$. Such a flux can be achieved at 1000 AU from a central protostar with $L_X \approx 10^{31}$ erg s$^{-1}$. The XDR abundances in the high-mass sources, however, require high X-ray fluxes ($F_X > 1$ erg s$^{-1}$ cm$^{-2}$), corresponding to luminosities $\gtrsim 10^{33}$ erg s$^{-1}$. X-rays can enhance the CN abundances up to three orders of magnitude compared to models without X-rays on scales of a few 100–1000 AU.

FUV fields produce CN abundances between $10^{-9}$–$10^{-8}$ (Fig. B.2) for temperatures $T \lesssim 300$ K and $G_0 = 5$–100. The abundances can be even higher for $T \gtrsim 300$ K at the relevant densities ($10^6$–$10^7$ cm$^{-3}$). Low FUV fields can thus enhance the CN abundances to similar values as high X-ray fluxes, if they can penetrate to large enough distances.

All models have $x(CN) < 10^{-10}$ for $n_H \gtrsim 10^6$ cm$^{-3}$ without the influence of X-rays or FUV fields. The observed constant fractional CN abundances for most sources, however, are $> 10^{-10}$ (Table 6). CN is therefore a clear indicator of enhanced X-rays or FUV fields.

## B.2. NO

Our observed NO abundances are of the order of $10^{-8}$ with respect to $H_2$ for the high-mass sources and $10^{-9}$ for the low-mass objects assuming constant fractional abundances (Table 6). In the jump models, the abundances of the high-mass sources are between $10^{-7}$ and a few times $10^{-6}$ (Tables 7 and 8). The observed NO column densities in the low and high-mass YSOs are $N(NO) \approx 10^{15}$–$10^{16}$ cm$^{-2}$.

The dependence of NO on the gas temperature, hydrogen density and X-ray flux is shown in Fig. B.3. The fractional NO abundance can get as high as $10^{-7}$ compared to total hydrogen in the models without X-rays for temperatures $300 \lesssim T(K) \lesssim 100$ and densities $n_H = 10^6$–$10^7$ cm$^{-3}$. The fractional NO abundance in models without X-rays is between a few times $10^{-10}$–$10^{-9}$ for $T \lesssim 100$ K and $n_H = 10^6$–$10^7$ cm$^{-3}$. X-rays can enhance these values by three orders of magnitude. The observed constant fractional abundances are comparable to the models without X-rays or low X-ray fluxes. The jump model abundances $x_{100K}$ (Table 7) are comparable to models with $F_X \approx 10^{-3}$–$1$ erg s$^{-1}$ cm$^{-2}$, corresponding to X-ray luminosities $\gtrsim 10^{31}$ erg s$^{-1}$. The XDR abundances require higher X-ray fluxes ($F_X \gtrsim 1$ erg s$^{-1}$ cm$^{-2}$).

Models with low FUV fields (Fig. B.4) show that the NO abundance is decreased to $\approx 10^{-12}$–$10^{-11}$ for temperatures $\lesssim 300$ K. NO is destroyed by photodissociation and in reactions with $C^+$. In the FUV models, NO is mainly produced in reactions of N with OH. OH is more abundant at high temperatures due to higher $H_2O$ abundances there, leading also to higher NO abundances for $T \gtrsim 300$ K in the FUV models. The observed NO abundances correspond to the model results with $G_0 \gtrsim 5$ and $T \gtrsim 300$ K.

The fact that chemical models without X-rays overestimate the NO abundance in the outer part of the envelope ($n_H \lesssim 10^6$ cm$^{-3}$) for most sources either suggests that FUV fields are present and reduce the overall NO abundance, that NO is frozen out on grains at lower temperatures or that some other – unknown – reduction mechanism exists for NO. Both X-rays and FUV, however, can produce the observed jump abundances.

## B.3. SO$^+$

Our observed SO$^+$ abundances in the high-mass objects are 5–$7 \times 10^{-11}$ assuming constant fractional abundances with respect to $H_2$ (Table 6). The $x_{100K}$ jump abundance is $2 \times 10^{-9}$ for AFGL 2591 and $9 \times 10^{-10}$ for W3 IRS5. The XDR jump abundance for AFGL 2591 is $3 \times 10^{-9}$ and $3 \times 10^{-8}$ for W3 IRS5. Inferred SO$^+$ column densities are $N(SO^+) \approx 10^{13}$ cm$^{-2}$.

The SO$^+$ abundance is studied as a function of the gas temperature, the X-ray flux and the hydrogen density (Fig. B.5). The temperature dependence of SO$^+$ is rather weak for $T \lesssim 300$ K. At higher temperatures, however, most oxygen is either in CO or $H_2O$, and SO$^+$ is less abundant. The upper $x_{XDR}$ limit derived for IRAS 16293–2422 is consistent with X-ray fluxes $F_X \gtrsim 1$ erg s$^{-1}$ cm$^{-2}$. Compared to the X-ray models for AFGL 2591 (Stäuber et al. 2005), much less SO$^+$ is produced. This is due to the much lower assumed initial sulphur abundance of only $10^{-8}$ for $T > 100$ K compared to $1.6 \times 10^{-6}$ for AFGL 2591. For $T < 100$ K the initial abundances are the same. The model results can thus be scaled by a factor of 160 for comparison with the high-mass sources at $T > 100$ K. High sulphur abundances ($x(S) \gtrsim 10^{-6}$), gas temperatures ($T \gtrsim 100$ K) and X-ray fluxes ($F_X \gtrsim 10^{-2}$ erg s$^{-1}$ cm$^{-2}$) are therefore required to explain the SO$^+$ observations towards the high-mass objects.

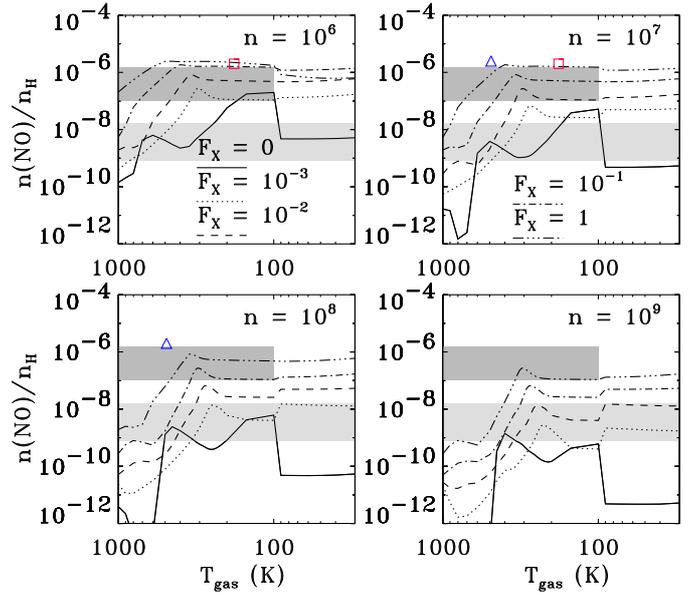

**Fig. B.3.** Modeled NO abundances as a function of the X-ray flux (erg s$^{-1}$ cm$^{-2}$), gas temperature and total hydrogen density (cm$^{-3}$). The light shaded region indicates the observed constant abundances. The dark shaded region indicates the range of $x_{100K}$. The square indicates $x_{XDR}$ for AFGL 2591 at the position of the average temperature and density in this region (Table 8); the triangle represents W3 IRS5.

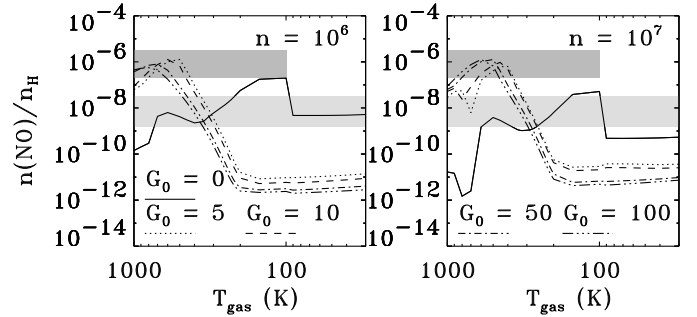

**Fig. B.4.** Modeled NO abundances as a function of the FUV radiation field $G_0$, gas temperature and total hydrogen density (cm$^{-3}$). The light shaded region indicates the observed constant abundances. The dark shaded region indicates the range of $x_{100K}$.

The FUV models in Fig. B.6 show that SO$^+$ is destroyed for temperatures lower than 300 K. In this temperature regime, the recombination of SO$^+$ is faster than its production. At higher temperatures, SO$^+$ can efficiently be produced in reactions of OH and $S^+$. The observed abundances of a few $\times 10^{-11}$ are only achieved at high temperatures ($T \gtrsim 300$ K). SO$^+$ is clearly enhanced either by X-rays or by FUV fields at high temperatures.

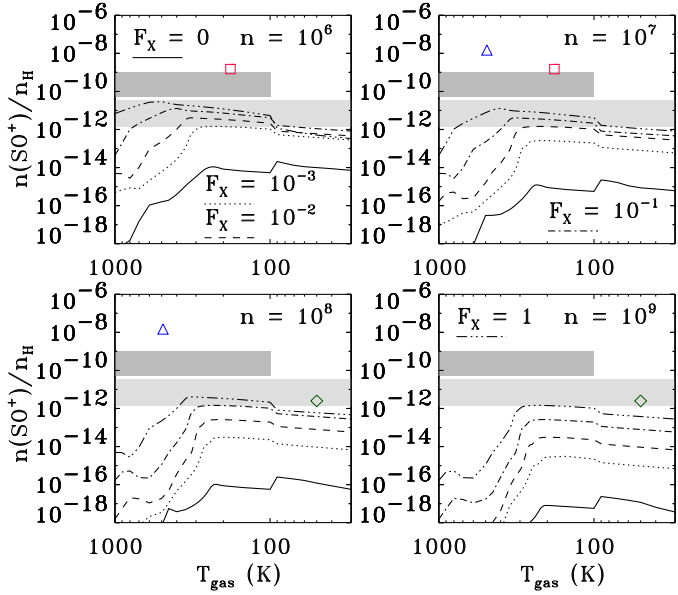

**Fig. B.5.** Modeled $SO^+$ abundances as a function of the X-ray flux (erg s$^{-1}$ cm$^{-2}$), gas temperature and total hydrogen density (cm$^{-3}$). The light shaded region indicates the observed constant abundances. The dark shaded region indicates the range of $x_{100K}$. The square indicates $x_{XDR}$ for AFGL 2591 at the position of the average temperature and density in this region (Table 8); the triangle represents W3 IRS5 and the diamond represents the upper limit in IRAS 16293–2422.

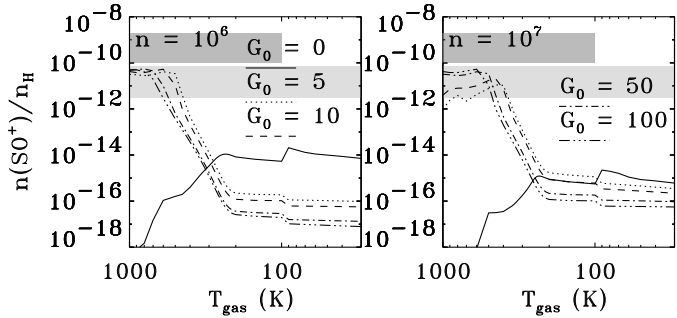

**Fig. B.6.** Modeled $SO^+$ abundances as a function of the FUV radiation field $G_0$, gas temperature and total hydrogen density (cm$^{-3}$). The light shaded region indicates the observed constant abundances. The dark shaded region indicates the range of $x_{100K}$.

### B.4. $CO^+$

$CO^+$ is observed with constant fractional $H_2$ abundances of $\approx 10^{-11}$–$10^{-10}$ in the high-mass objects and $1.5 \times 10^{-12}$ in IRAS 16293–2422 (Table 6). Beam averaged column densities are $\approx 10^{12}$ cm$^{-2}$ for the high-mass objects and a few $\times 10^{11}$ cm$^{-2}$ for the low-mass YSOs, depending on the excitation temperature (Table A.1). The $x_{100K}$ jump abundances are $\approx 10^{-10}$–$10^{-9}$ for $T \gtrsim 100$ K (Table 7) and $x_{XDR}(CO^+) \approx 10^{-12}$–$10^{-9}$ (Table 8).

It is well-known, that chemical models usually have difficulties in producing the observed $CO^+$ column densities (e.g., Fuente et al. 2006). However, to study the general conditions in the gas to produce the observed $CO^+$ abundances, they are calculated as functions of the X-ray flux, hydrogen density and gas temperature (Fig. B.7). The $CO^+$ abundances are fairly constant with temperature for $T \lesssim 300$ K. At higher temperatures, most oxygen is driven into $H_2O$ even for high X-ray fluxes (Stäuber et al. 2006) and the $CO^+$ abundance decreases. The observed constant fractional abundances of the order $10^{-12}$–$10^{-11}$ are reached for densities $n_H \lesssim 10^7$ cm$^{-3}$ and X-ray fluxes $F_X \gtrsim 10^{-1}$ erg s$^{-1}$ cm$^{-2}$. The XDR jump abundances require X-ray fluxes exceeding $F_X = 1$ erg s$^{-1}$ cm$^{-2}$, corresponding to $L_X \gtrsim 10^{33}$ erg s$^{-1}$.

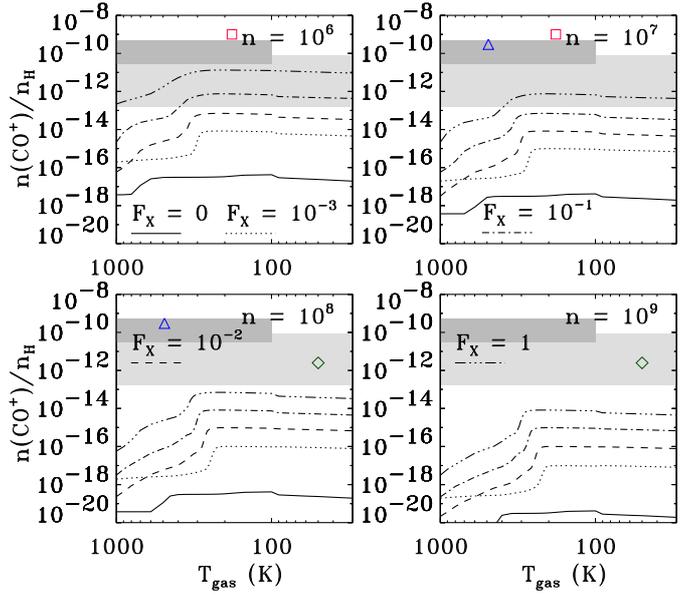

**Fig. B.7.** Modeled $CO^+$ abundances as a function of the X-ray flux (erg s$^{-1}$ cm$^{-2}$), gas temperature and total hydrogen density (cm$^{-3}$). The light shaded region indicates the observed constant abundances. The dark shaded region indicates the range of $x_{100K}$. The square indicates $x_{XDR}$ for AFGL 2591 at the position of the average temperature and density in this region (Table 8); the triangle represents W3 IRS5 and the diamond represents IRAS 16293–2422.

Figure B.8 shows how the $CO^+$ abundance depends on the FUV field strength. The fractional abundances are two orders of magnitude below the observed values for $T \lesssim 300$ K. The fractional abundances can increase to $x(CO^+) \approx 10^{-12}$–$10^{-10}$ with respect to total hydrogen for gas temperatures $\gtrsim 300$ K. $CO^+$ is efficiently produced in reactions of $C^+$ and OH with the latter being more abundant at higher temperatures due to dissociations of $H_2O$ (see also Stäuber et al. 2004). The $CO^+$ observations can thus be interpreted with either a high X-ray flux ($F_X > 1$ erg s$^{-1}$ cm$^{-2}$) or an FUV flux ($G_0 \gtrsim 5$) at high gas temperatures.

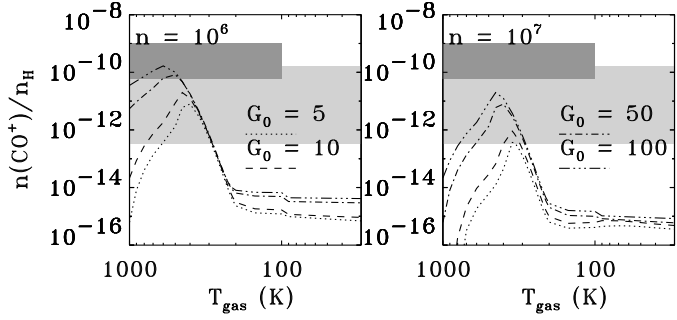

**Fig. B.8.** Modeled CO$^+$ abundances as a function of the FUV radiation field $G_0$, gas temperature and total hydrogen density (cm$^{-3}$). The light shaded region indicates the observed constant abundances. The dark shaded region indicates the range of $x_{100K}$.

**Table B.2.** Summary of the results from the general parameter study[a].

| Molecule | X-rays ($F_X$) erg s$^{-1}$ cm$^{-2}$ | FUV ($G_0$) $T \lesssim 300$ K | FUV ($G_0$) $T \gtrsim 300$ K |
|---|---|---|---|
| CN (const.) | $\gtrsim 10^{-4}$ | $\gtrsim 5$ | $\gtrsim 5$ |
| CN (jump) | $\gtrsim 10^{-3}$ | $\gtrsim 5$ | $\gtrsim 5$ |
| NO (const.) | $0$–$10^{-2}$ | - | $\gtrsim 5$ |
| NO (jump) | $\gtrsim 10^{-3}$ | - | $\gtrsim 5$ |
| SO$^+$ (const.) | $\gtrsim 10^{-3}$ | - | $\gtrsim 5$ |
| SO$^+$ (jump) | $\gtrsim 10^{-2}$ | - | $\gtrsim 5$ |
| CO$^+$ (const.) | $\gtrsim 10^{-1}$ | - | $\gtrsim 5$ |
| CO$^+$ (jump) | $> 1$ | - | $\gtrsim 5$ |

[a] The lower limit or range of X-ray and FUV fluxes for each molecule corresponds to the values that either fit the observed constant fractional abundances or the jump abundances ($x_{100K}$, $x_{XDR}$). NO, SO$^+$ and CO$^+$ cannot be modeled with $G_0 \gtrsim 5$ and $T \lesssim 300$ K.